%% file: top.tex
\documentclass[journal]{IEEEtran}
\usepackage{cite}
\usepackage{amsmath,amssymb,amsfonts}
\usepackage{algorithmic}
\usepackage{graphicx}
\usepackage{textcomp}
\usepackage{xcolor}
\usepackage[]{hyperref}
\usepackage{subfigure}
\usepackage{booktabs}
\usepackage{multirow}
\usepackage{siunitx}
\usepackage{algorithm}
\usepackage{CJK}
\usepackage{dblfloatfix}

\ifCLASSINFOpdf
\else
\fi

\begin{document}

%
\title{Taming Process Variations in CNFET for Efficient Last Level Cache Design}
%
%
%

\author{Dawen Xu,
        Zhuangyu Feng,
        Cheng Liu, \IEEEmembership{Member, IEEE}
        Li Li,
        Ying Wang, \IEEEmembership{Member, IEEE} \\
        Yuanqing Cheng, \IEEEmembership{Senior Member, IEEE}
        Huawei Li, \IEEEmembership{Senior Member, IEEE}
        and Xiaowei Li \IEEEmembership{Senior Member, IEEE}

\thanks{Cheng Liu is the corresponding author.} 
\thanks{This article was presented in part at Proceedings of the 24th Asia and South Pacific Design Automation Conference, 2019.} 

\thanks{Dawen Xu, Zhuangyu Feng, and Li Li are with both Hefei University of Technology, Hefei 230009, Anhui, China and SKLCA, Institute of Computing Technology, Chinese Academy of Sciences, Beijing 100190, China.}

\thanks{Cheng Liu, Ying Wang, and Xiaowei Li are with SKLCA, Institute of Computing Technology, Chinese Academy of Sciences, Beijing 100190, China. (e-mail: liucheng@ict.ac.cn)}

\thanks{Huawei Li is with both SKLCA, Institute of Computing Technology, Chinese Academy of Sciences, Beijing 100180, China and Peng Cheng Laboratory, Shenzhen, 518055, China.}

\thanks{Yuanqing Cheng is with the School of Microelectronics, Beihang University, Beijing 100191, China.}
}

%
%

\markboth{IEEE Transactions on Very Large Scale Integration (VLSI) Systems }%
{Shell \MakeLowercase{\textit{et al.}}: Bare Demo of IEEEtran.cls for Journals}
%


\maketitle

\begin{abstract}
Carbon nanotube field-effect transistors (CNFET) emerge as a promising alternative to CMOS transistors for the much higher speed and energy efficiency, which makes the technology particularly suitable for building the energy-hungry last level cache (LLC). However, the process variations (PVs) in CNFET caused by the imperfect fabrication lead to large timing variation and the worst-case timing dramatically limits the LLC operation speed. Particularly, we observe that the CNFET-based cache latency distribution is closely related to the LLC layouts. For the two typical LLC layouts that have the CNT growth direction aligned to the cache way direction and cache set direction respectively, we proposed variation-aware set aligned (VASA) cache and variation-aware way aligned (VAWA) cache in combination with corresponding cache optimizations such as data shuffling and page mapping to enable low-latency cache for frequently used data. According to our experiments, the optimized LLC reduces the average access latency by 32\% and 45\% compared to the baseline designs on the two different CNFET layouts respectively while it improves the overall performance by 6\% and 9\% and reduces the energy consumption by 4\% and 8\% respectively. In addition, with both the architecture induced latency variation and PV incurred latency variation considered in a unified model, we extended the VAWA and VASA cache design for the CNFET-based NUCA and the proposed NUCA achieves both significant performance improvement and energy saving compared to the straightforward variation-aware NUCA.

\end{abstract}

\begin{IEEEkeywords}
Process Variation, Last Level Cache, CNFET, Variation-aware Cache
\end{IEEEkeywords}

\IEEEpeerreviewmaketitle

\input{intro}
\input{background}

\input{motivation}
\input{vasa}
\input{vawa}
\input{NUCA}

\input{experiment}
\section{Conclusion} \label{sec:conclusion}
In this paper, we investigated the influence of CNT density variations on the LLC access latency and observed that LLC access latency shows large variations for both a set aligned layout and a way aligned layout. The worst-case timing dramatically limits the LLC operation speed when the entire LLC works at a unified clock. To address the problem, we proposed a variation-aware cache design to enable optimized operation speed for different parts of the cache with distinct timing. In addition, we take the memory access preference of the applications into consideration and have the frequently used data placed at cache parts with lower access latency. For the CNFET LLC with set aligned layout that has distinct access latency across the different cache ways, we propose a data shuffling strategy to have the frequently used data swapped to the cache ways with lower latency at runtime. For the CNFET LLC with way aligned layout that has distinct access latency across the different cache sets, we propose a page mapping strategy to have the frequently used data mapped to cache sets with lower latency. According to our experiments on a set of representative benchmark, the optimized CNFET-based LLC improves the average performance by 6\% and 9\% for the two different layouts respectively compared to the baseline designs while it reduces the energy consumption by 4\% and 8\% respectively. Moreover, we extended the proposed cache design for an NUCA that has both architecture-inherent latency variations and CNFET PV incurred latency variations considered with a unified page mapping optimization. The proposed CNFET-based NUCA optimization further achieves significant performance speedup and energy reduction compared to the basic variation-aware cache designs.





\ifCLASSOPTIONcaptionsoff
  \newpage
\fi



%
\bibliographystyle{IEEEtran}
\bibliography{top}

\end{document}

%% file: intro.tex
\section{Introduction} \label{sec:intro}
\IEEEPARstart{C}{ache} especially the last-level cache (LLC), which bridges the ever-increasing gap between fast processing units and the relatively slow main memory, is critical to both the performance and energy efficiency of processors \cite{zang2013survey}. Although a large cache is preferred for higher cache hit rate in most applications, the cache operation speed and energy efficiency can deteriorate substantially. The contradiction between the cache size and the cache operation speed as well as the cache energy consumption is essentially attributed to inherent circuit architectures. Thus, it is rather difficult to address with the existing CMOS technology. Compared to CMOS transistors, emerging carbon nanotube field-effect transistors (CNFET) exhibit much larger 'on' current and lower 'off' leakage current \cite{qiu2017scaling} \cite{qiu2017limit}. Since the large 'on' current ensures high operation speed and the low 'off' leakage current indicates ultra-low static power consumption, CNFET technology becomes promising for the high-capacity, high-performance, and energy-efficient cache designs. According to the work in \cite{carbon2017ISLPED, hills2018understanding, shulaker2013carbon, bishop2020fabrication}, the energy efficiency of CNFET-based processors can be an order of magnitude higher than that of the CMOS-based processors at even sub-10 \SI{}{\nano\meter} technology. The great advantages of CNFET-based integrated circuits attract growing attentions of researchers recently \cite{qiu2017limit} \cite{bishop2020fabrication} \cite{qiu2017scaling} \cite{exploring2019aspdac} \cite{xu2020cnt}.

Despite the promising advantages, CNFET technology that relies on chemical synthesis still suffers the imprecise control of the carbon nanotube (CNT) growth and the imperfect fabrication can result in dramatic CNFET process variations (PVs) \cite{analysis2020TVLSI, patil2010scalable, lin2009metallic, patil2007automated, grigoleit2011cntfet, sheikh2016cnfet, almudever2015variability, zhang2009carbon, zhang2010carbon, zhang2011overcoming, zhang2009probabilistic}. There are many different types of CNFET PVs such as the metallic CNT (m-CNT) \cite{zhang2009probabilistic, lin2009metallic, patil2010scalable, patil2007automated, zhang2010carbon, grigoleit2011cntfet}, the CNT density variation \cite{zhang2009carbon,zhang2010carbon, zhang2011overcoming, zhang2011characterization}, the mis-positioned CNTs \cite{zhang2010carbon, patil2010scalable, patil2007automated}, and the CNT diameter variation \cite{zhang2010carbon}. They can lead to CNT drive current variation and circuit delay variation eventually. According to the work in \cite{li2018cnfet}, the delay of a CNFET-based ALU ranges from 4 to 8 cycles and the delay of a register file can be 3$\times$ higher. While LLC includes a large number of CNFETs, the PV-induced circuit latency variation can be large. If the CNFET-based LLC design is applied directly, it can only work at the worst timing and the cache performance will be substantially limited by the PVs.

In order to address the above problem, notable efforts have been devoted to the CNFET-based architectural design and exploration. A micro-architectural timing model was proposed in \cite{li2015microarchitectural} to evaluate the delay distribution of different CNFET layouts. On top of the models, Li Jiang et al. in \cite{jiang2017cnfet} \cite{li2016cnfet} explored the CNFET-based design of modern SIMD register files and proposed a novel register assignment method to take advantage of the delay variation caused by the asymmetrically correlated variation. Patil et al. \cite{patil2010scalable} proposed a VLSI-compatible metallic CNT removal (VMR) approach to create basic CNFET circuits such as half-adders and D-latches immune to CNT imperfections. Zhang et al. \cite{zhang2010carbon} proposed to upsize all the CNFETs and cover the variations with additional chip area and energy consumption. Patil et al. \cite{hills2018trig} utilized an iterative gray code (TRIG) to reduce the timing errors caused by the process variations for a binary neural network accelerator. This approach in combination with the inherent fault-tolerance of neural networks achieves competitive performance and energy efficiency with minor model accuracy loss. Nevertheless, there are still a lack of approaches that can be efficiently applied to the large and regular LLC design from the perspective of architectural design.

CNT density variation is a major source of CNFET imperfection and we particularly focus on the variation caused by correlation of CNT count in CNFETs in this work. According to \cite{zhang2011characterization} \cite{li2015microarchitectural}, the spatial correlation of CNT count for the aligned CNT growth direction is strongly asymmetric (direction-dependent). CNFETs with the CNT growth direction aligned are similar to each other while CNFETs with the CNT growth direction unaligned suffer larger variations. Based on this feature, we investigated the CNFET-based LLC design under two different CNT growth directions and drew the following observations. When the CNT growth direction is parallel to the bitline in LLC, the cache access latency exhibits similar latency across the different cache sets and we name this pattern as set aligned cache. When the CNT growth direction is parallel to the wordline in LLC, the cache access latency will be close to each other across the different cache ways and we name it as way aligned cache.

Accordingly, we further develop corresponding variation-aware cache structures to adapt to the two distinct cache latency variations. For the set aligned cache and way aligned cache, we propose a variation-aware data shuffling strategy and a variation-aware page mapping strategy to have the frequently used data placed at the faster cache ways and cache sets respectively such that different parts of the cache with distinct timing can work at their optimized performance and the potential of the CNFET-based LLC with variable access latency can be unleashed. Finally, we adapt the proposed variation-aware cache architectures to a CNFET-based NUCA architecture. By combining both the architecture induced latency variation and PV incurred latency variation in a unified latency model, we can further enhance the CNFET-based NUCA performance through a unified page mapping. According to our experiments, the proposed cache optimizations show significant superiority compared to a straightforward variation-aware NUCA. The contributions of this work can be summarized as follows:

\begin{itemize}
\item We investigated the CNFET-based cache latency variation caused by typical direction dependent CNT layouts and proposed variation-aware set aligned (VASA) cache and variation-aware way aligned (VAWA) cache respectively to make best use of the CNFET cache architecture with variable latency.

\item For the set aligned cache layout, we developed a latency-aware data shuffling strategy to assign the most recent data to the fast cache ways. This strategy reduces the LLC access time by 32\% on average, which improves the average performance by 6\% and saves the energy consumption by 4\% on average compared to the worst-case timing design. For the way aligned cache, we proposed a variation-aware page mapping to ensure that the most frequent used data are mapped to fast cache sets. It reduces the LLC access time by 45\%, improves the average performance by 9\% and reduces the energy consumption by 8\% compared to the baseline design.

\item We built an NUCA on top of VAWA cache and VASA cache respectively, and exploited the NUCA potential by considering both the NUCA architecture induced latency variation and PV incurred latency variation in a unified model. According to our experiments, the proposed CNFET-based NUCA achieves both significant performance improvement and energy saving on top of the straightforward VASA and VAWA NUCA.
\end{itemize}

The rest of this paper is organized as follows. In Section \ref{sec:background}, we brief CNFET variations and previous variation-aware circuit designs. In Section \ref{sec:motivation}, we discuss the delay variation distribution of the CNFET-based cache with different layouts and motivate this work. In Section \ref{sec:vawa}, we mainly detail the proposed variation-aware way aligned (VAWA) cache architecture in which the cache access latency varies in different cache ways. In Section \ref{sec:vasa}, we mainly detail the proposed variation-aware set aligned (VASA) cache architecture in which the cache latency varies in different cache sets. In Section \ref{sec:NUCA}, we further investigate the use of VASA and VAWA on an NUCA architecture. In Section \ref{sec:experiment}, we show the experimental results of the proposed variation-aware cache architectures. In Sections \ref{sec:conclusion}, we conclude this paper.

%% file: background.tex
\section{Background and Related Work}\label{sec:background}
\begin{figure}
	\center{\includegraphics[width=0.8\linewidth]{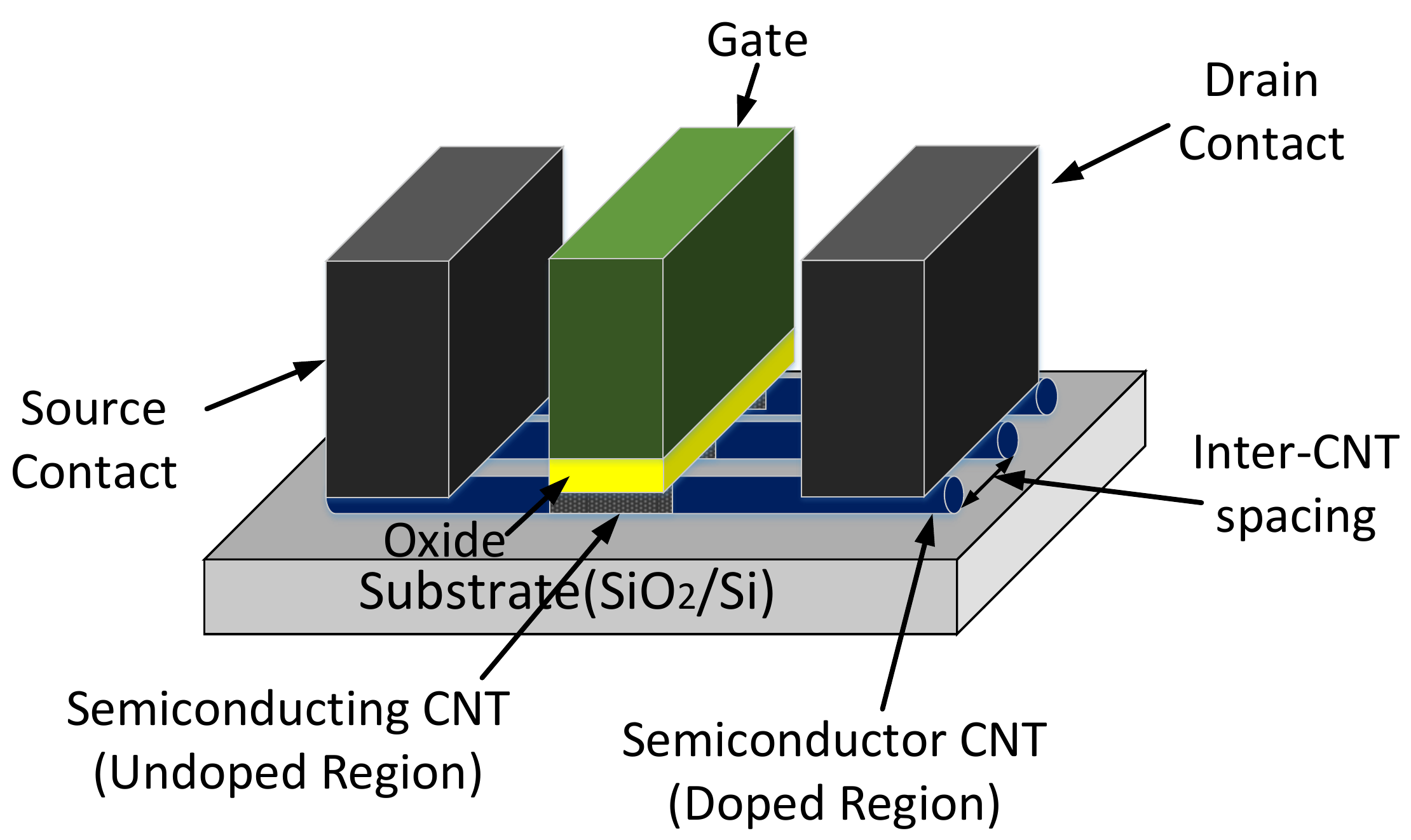}}
	\vspace{-0.5em}
    \caption{Basic CNFET Structure.}
\label{fig:CNFET-struct}
\vspace{-1em}
\end{figure}

\begin{figure}
	\center{\includegraphics[width=0.6\linewidth]{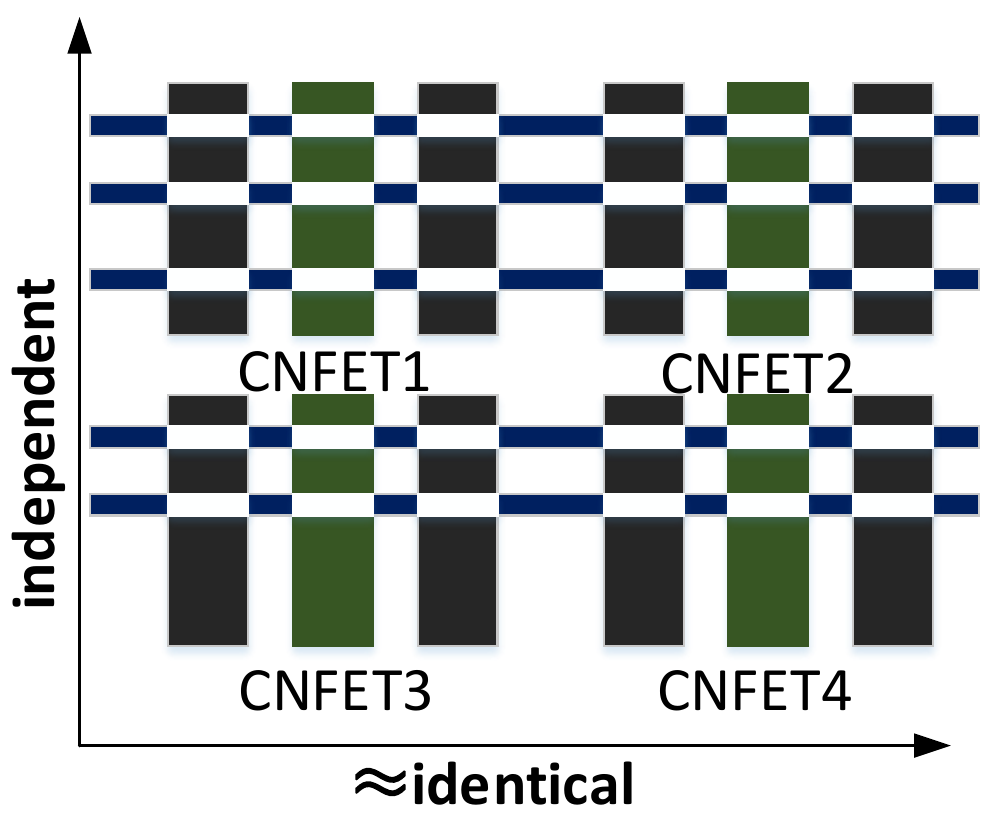}}
	\vspace{-0.5em}
    \caption{Correlation between CNFETs at different locations.}
\label{fig:CNFET-correlation}
\vspace{-1em}
\end{figure}

\begin{figure*}
	\center{\includegraphics[width=0.95\linewidth]{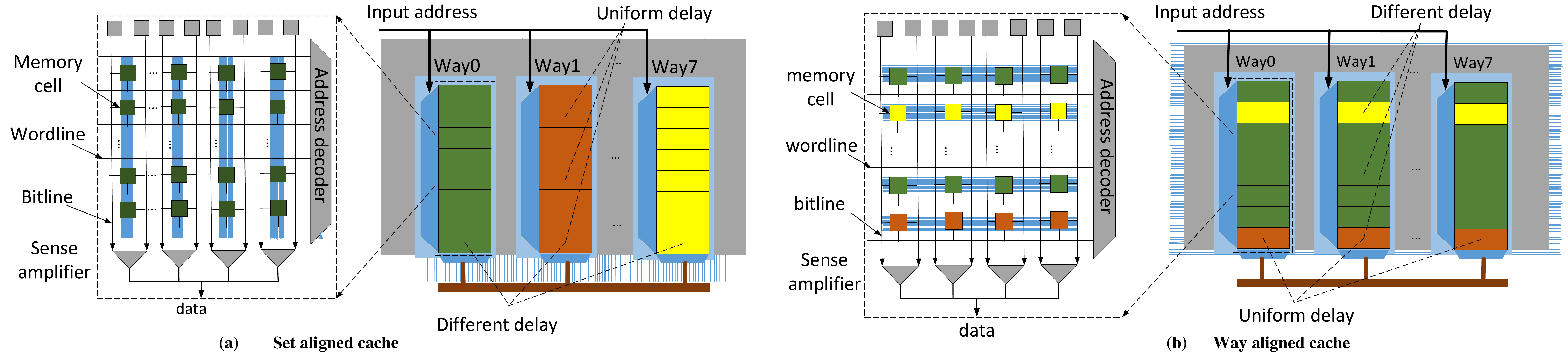}}
	\vspace{-0.5em}
    \caption{Cache architecture with different CNT layouts. When the CNT growth direction is parallel to the bitline, we call it set aligned cache. When the CNT growth direction is parallel to the wordline, we call it way aligned cache.}
\label{fig:CNFET-layouts}
\vspace{-1em}
\end{figure*}

\subsection{CNT Density Variation}
CNTs act as channels modulated by the transistor gate as shown in Figure \ref{fig:CNFET-struct}. The regions of the CNTs under the CNFET gate are undoped while the CNFET source and drain regions of the CNTs are heavily doped. Each CNFET usually contains multiple CNTs between source to drain contacts. Unlike the gates, source, drain, and interconnects that are defined by conventional lithography, the inter-CNT spacing is controlled by the CNT growth process. Due to the lack of precise control over CNT positioning during chemical synthesis, significant variations can exist in the space between CNTs. Accordingly, the CNT density i.e. CNT count per unit width of CNFET is non-uniform. While multiple CNTs can conduct current in parallel to supply sufficient drive current, CNFET fabricated in the presence of CNT density variations will cause large variation in its drive current and may even result in significant probability of complete failure. CNT variations have become a major hurdle to the design of large-scale CNFET circuits \cite{zhang2011characterization}. 

Particularly, the authors in \cite{zhang2011characterization} \cite{li2015microarchitectural} noticed that CNFETs with aligned CNTs are likely to have similar CNT count and thus similar electrical properties to each other \cite{zhang2010carbon}. While CNTs in CNFETs typically are aligned over long distances ($>100$ \SI{}{\micro\meter}), these CNFETs aligned to the same CNT growth direction have little variation among them. An example of CNFETs with CNT variation is shown in Figure \ref{fig:CNFET-correlation}. When the CNT growth direction is horizontal, both channels of CNFET1 and CNFET2 are aligned with the same direction. In this case, the drive current of both CNFET1 and CNFET2 is close to each other. In contrast, CNFET1 and CNFET3 are quite different in terms of CNT density. Their drive current and circuit latency varies accordingly.

\subsection{Variation-tolerant Design Approaches}
To address the CNT variation problem, many approaches from distinct angles have been proposed. Albert et al. proposed to make full use of the asymmetric correlation of CNFETs fabricated on aligned carbon nanotubes and combine an array of m-CNT-susceptible CNFETs to achieve high $I_{on}$/$I_{off}$ and high current drive \cite{lin2009metallic}, which greatly reduces the negative influence of m-CNTs. Matthias et al. proposed a heuristic placement to optimize the circuit layout such that the critical path of the circuit can be placed on CNFETs with lower delay by taking advantage of the CNT asymmetric correlation \cite{beste2014layout}. Feng et al. developed a novel test strategy which conducts the correlated faults induced by m-CNTs before the march test to achieve high fault coverage and reduce the CNFET-based SRAM test cost \cite{xie2015jump, li2016novel}. Jie et al. developed a parameterized model for the CNT density variations and further discussed the CNFET layout design optimizations on top of the model, which can significantly improve the reliability and yield of the CNFET-based design \cite{zhang2009carbon} \cite{zhang2011characterization}.

Different from the above works that mainly address the CNT variations at transistor or gate level design, many researchers address the CNT variations with novel architectural designs. Hao et al. proposed to reduce the transistor-level redundancy of CNFETs for more energy-efficient and faster cores but lower reliability \cite{wang2013improving}. With the same resources and energy constraints, this approach that combines a many-core architecture and a core-level redundancy achieves significantly higher performance and energy efficiency compared to the many-core architecture with only transistor-level redundancy. Gage et al. proposed to utilize an iterative gray code to overcome the errors induced by the process variations in the CNFET-based multiply-accumulate (MAC) circuits \cite{hills2018trig}. On top of the fault-tolerant MAC, binary neural network accelerators targeting at lightweight inference can be built to overcome the CNT variations. Jiang et al. investigated the influence of CNFET variations on GPU register file (RF) and proposed a novel RF structure and register alignment technique that can tolerate the CNFET density variations and achieve higher performance at the same time \cite{li2018cnfet}. Particularly, they demonstrated that the performance of RF with a proper structure design can be 53\% higher than that conformed to the worst timing. CNFET-based LLC with a large number of SRAM cells is also dramatically affected by the CNT density variations and novel cache structure is highly demanded to make good use of the CNFET technology for both higher performance and energy efficiency.

%% file: motivation.tex
\section{Motivation} \label{sec:motivation}
In this section, we mainly explore the influence of CNT density variations on the access latency of the CNFET-based cache. As the CNT density variation is closely related with the CNT growth direction, the CNFET-based cache access latency variation depends on the CNT layouts accordingly. The two typical CNT layouts i.e. the set aligned layout and the way aligned layout are presented in Figure \ref{fig:CNFET-layouts}. The set aligned cache as shown in Figure \ref{fig:CNFET-layouts}(a) has similar cache access latency among the different cache sets but variable access latency across the different cache ways. In contrast, the way aligned cache as shown in Figure \ref{fig:CNFET-layouts}(b) has similar cache access latency across the different cache ways but variable access latency among the different cache sets or cache lines. 

\begin{figure}[t]
	\center{\includegraphics[width=0.9\linewidth]{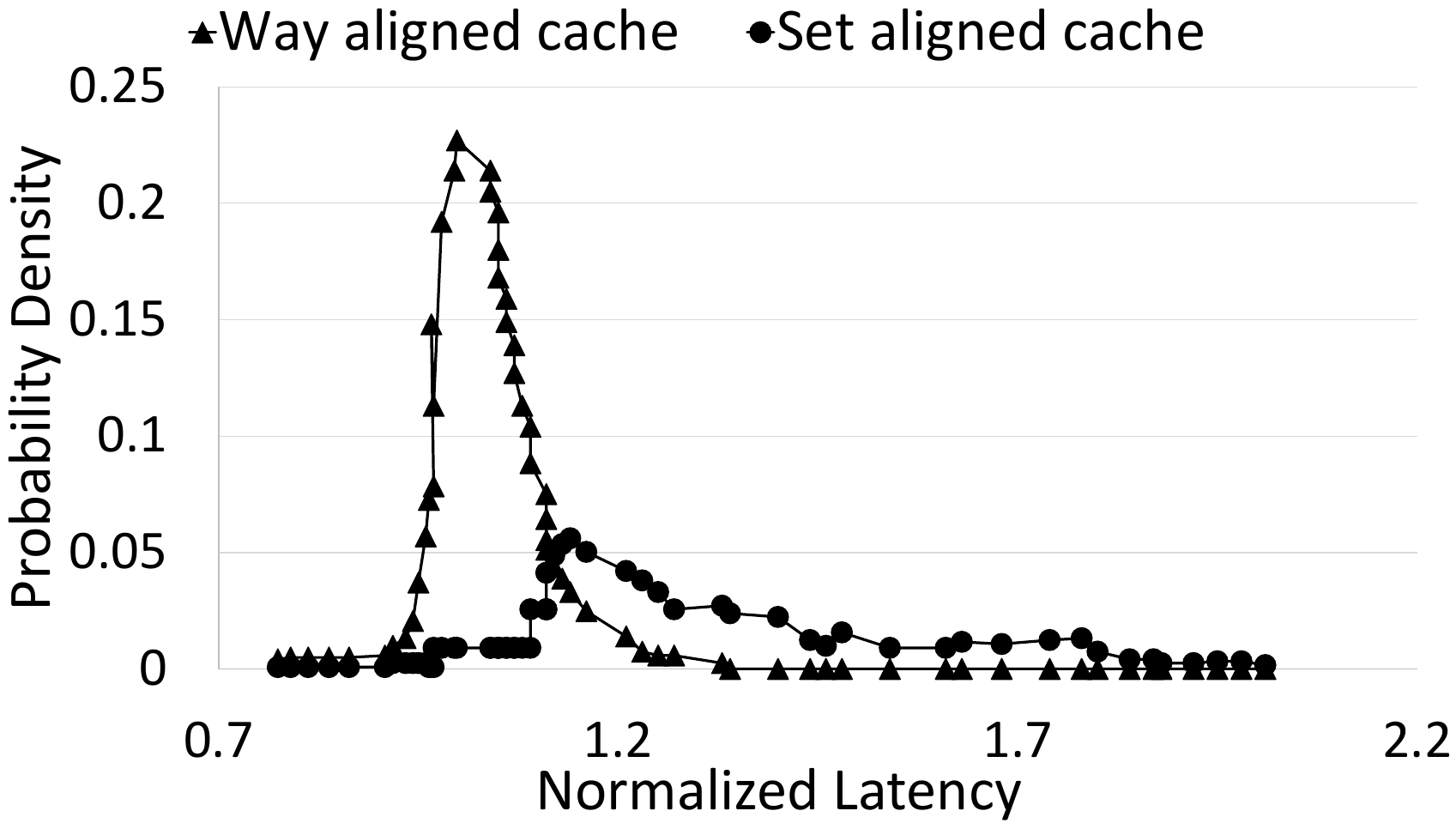}}
    \caption{CNFET-based cache latency distribution with different CNT layouts.}
\label{fig:lat-dist}
\vspace{-0.5em}
\end{figure}

To gain insight of the cache latency variations, we analyzed the cache latency distribution based on a CNT-count model proposed in \cite{zhang2011characterization} and a cache micro-architecture timing model proposed in \cite{li2015microarchitectural}. The resulting CNFET-based cache latency distribution is presented in Figure \ref{fig:lat-dist}. The cache access latency is normalized to that of an ideal design without any CNT density variations. The relevant CNT parameters used in the model is detailed in Table \ref{tab:cnt}. It can be observed that the cache latency varies in a broad range and the worst cache latency can be $3\times$ larger than the best case. Meanwhile, it can be found that the latency of the way aligned cache exhibits relatively narrower range and larger variation compared to the set aligned cache. This is mainly because that the pass transistors in the same wordline exhibit highly correlated gate capacitance and the load capacitance may be either very large or extremely small due to the asymmetric spatial correlation for the way aligned cache\cite{li2015microarchitectural}.

\begin{table}
\centering
\caption{CNT parameters}
\begin{tabular}{ll} 
\hline
Parameters                                              & Value                 \\ 
\hline
Cache size                                              & 2MB \\
Cache way                                               & 8 \\
Gate width                                              & \SI{32}{\nano\meter}\\
CNT count distribution: \\mean $\mu$, standard deviation $\delta$~ & $\mu=9$, $\delta=2.1$ \cite{zhang2011characterization} \\
Probability of removed m-CNT                            & $P_{rm}=99.9\%$ \cite{zhang2011characterization}  \\
Probability of m-CNT                                    & $P_{m}=5\%$ \cite{zhang2011characterization}  \\
Probability of aligned                                  & $P_{align}=5\%$ \cite{patil2009vmr}  \\
Probability of removed s-CNT                            & $P_{rs}=5\%$ \cite{zhang2011characterization}    \\
\hline
\end{tabular}
\label{tab:cnt}
\end{table}

\begin{figure}[t]
	\center{\includegraphics[width=0.9\linewidth]{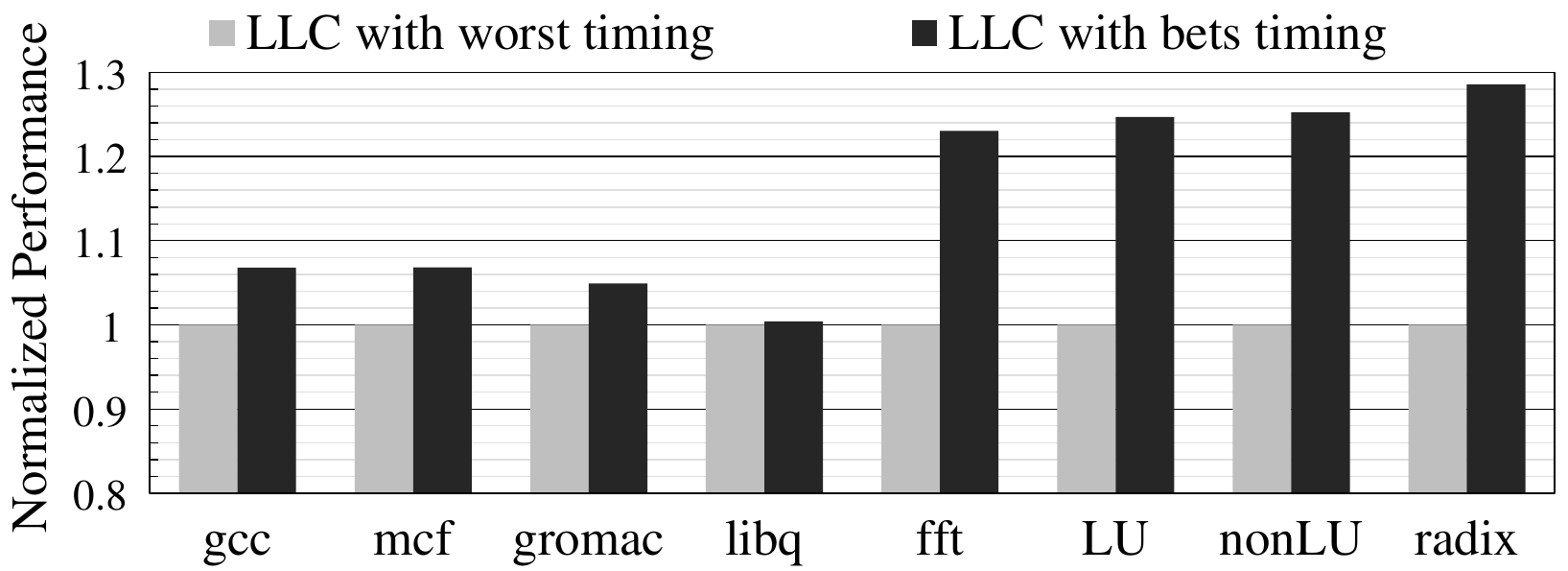}}
    \caption{Normalized performance of processors with the CNFET-based cache running at the best timing (6-cycle) and the worst timing (12-cycle).}
\label{fig:lat-perf}
\vspace{-0.5em}
\end{figure}

In addition, we configured the CNFET-based LLC with the best timing and the worst timing respectively and applied them to a general purposed processor. Then we evaluated the performance under a set of typical applications selected from Spec CPU2006 and Splash2 \cite{woo1995splash}. As shown in \autoref{fig:lat-perf}, the performance of the processor equipped with the LLC with the best timing is much higher than that with the worst timing. It also demonstrates that the access latency of the LLC is of vital importance to the performance of these applications in general though some of the applications like libquantum are not sensitive to the access latency. While we c
an not have the entire LLC working at the best timing in practice directly because the timing violation can corrupt the data stored in it, intensive optimizations are required to exploit the potential of the CNFET-based LLC.

%% file: vasa.tex
\section{Variation-aware Set Aligned (VASA) Cache} \label{sec:vasa}
If we have the entire CNFET-based cache to work at a fixed clock like conventional cache, the cache operation speed will be limited to the worst timing, which is a waste to the regions of the cache that can potentially operate at higher speed. To address this problem, we propose a variation-aware cache architecture that allows the cache to operate at variable speed and adapts to latency variations of the different cache regions. As the cache access latency variation patterns are closely related with the underlying CNFET-based cache layouts, the cache implementations with two typical cache layouts i.e. the set aligned cache and the way aligned cache exhibit distinct latency variations. Both of them will be illustrated in this work and we focus on the variation-aware set aligned (VASA) cache design and optimization in this section.

\begin{figure}
	\center{\includegraphics[width=0.95\linewidth]{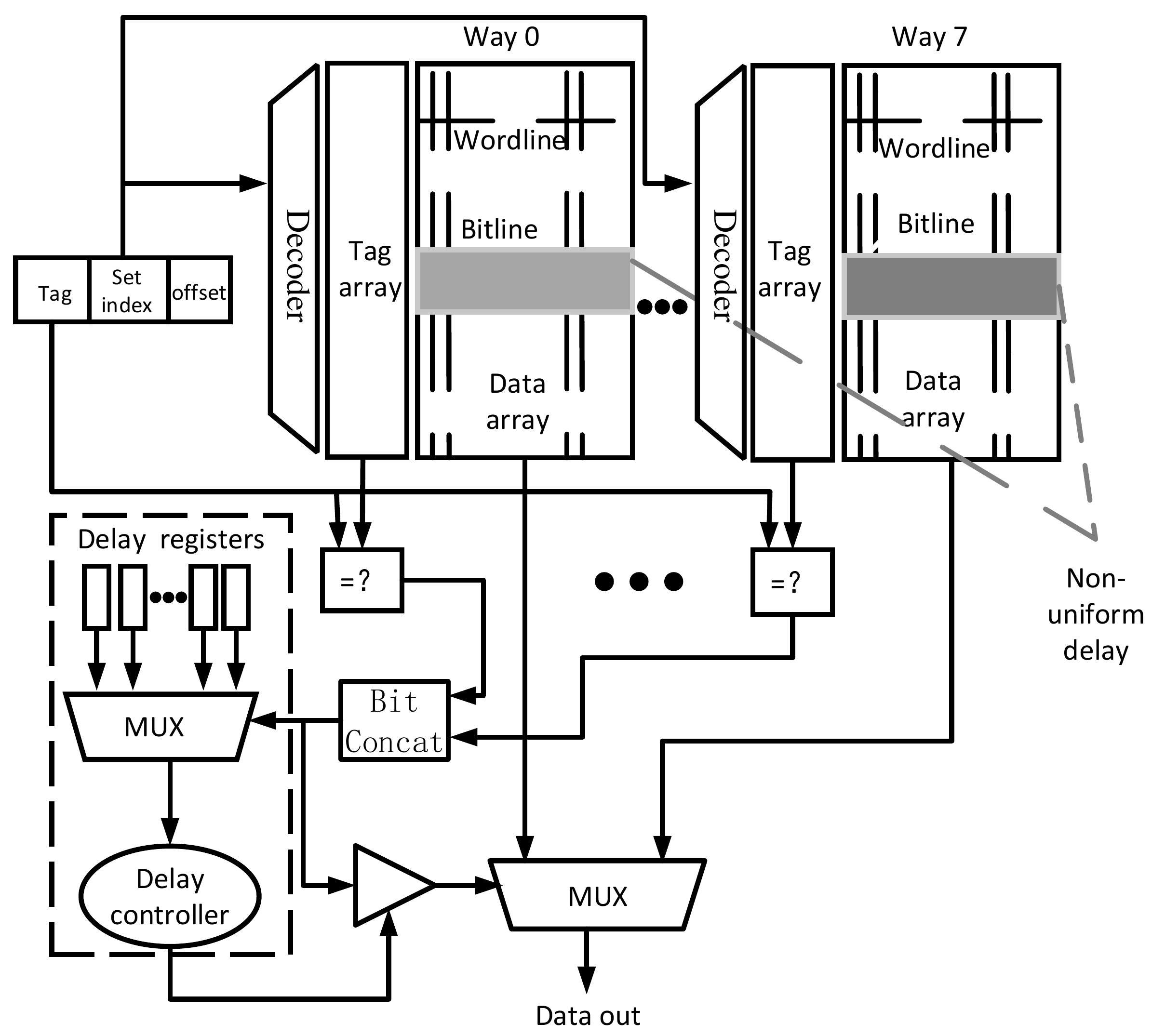}}
	\vspace{-0.5em}
    \caption{Variation-aware set aligned cache architecture.}
\label{fig:vasa-arch}
\vspace{-1em}
\end{figure}

\begin{figure*}
	\center{\includegraphics[width=0.85\linewidth]{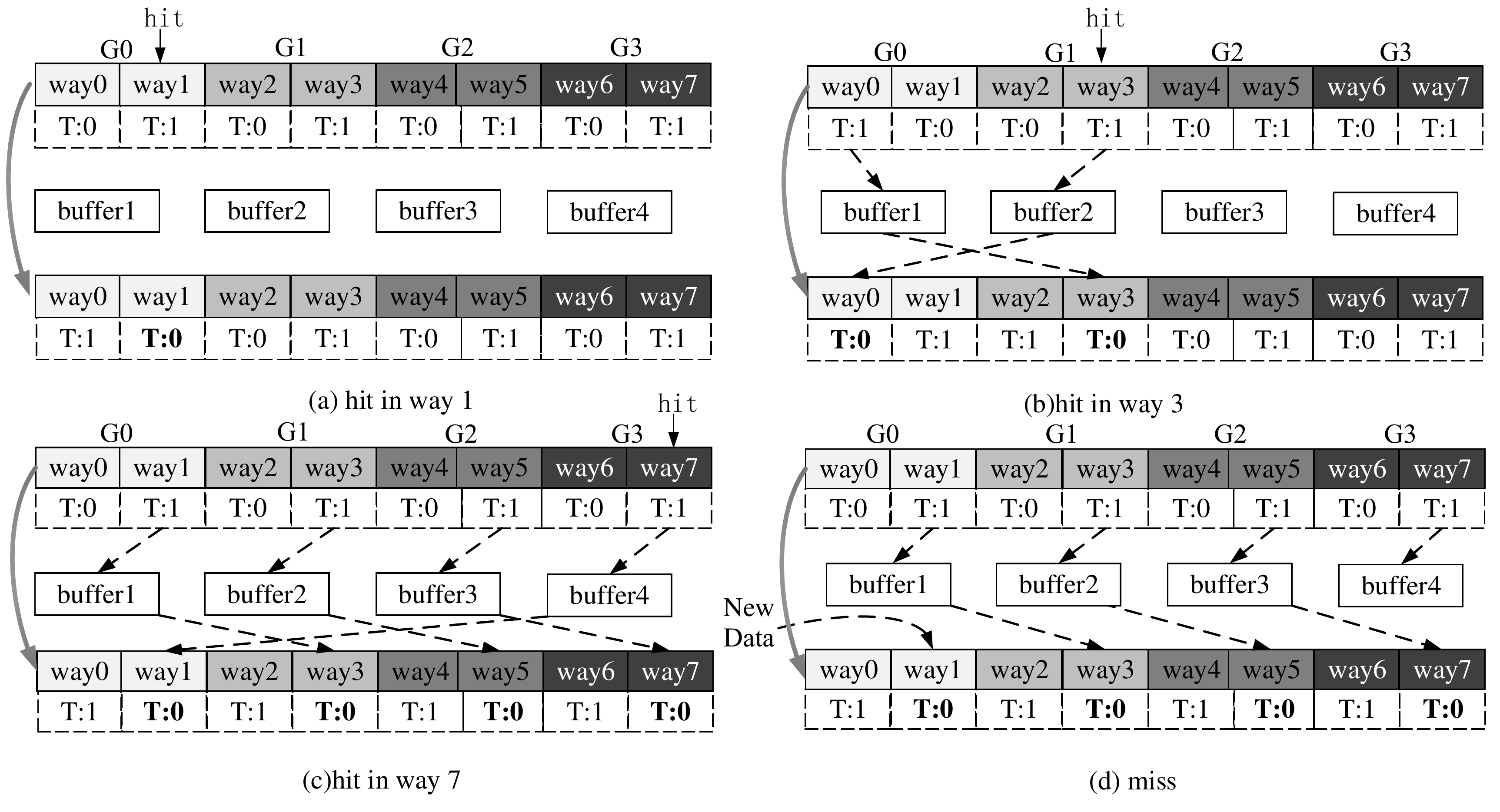}}
	\vspace{-0.5em}
    \caption{Examples of the proposed latency-aware data shuffling under four different scenarios.}
\label{fig:data-shuffling}
\vspace{-1em}
\end{figure*}

\subsection{VASA Cache Architecture}
For the set aligned cache, the access latency of cache sets in the same cache bank is similar but the access latency of the different cache ways varies. To unleash the performance potential of the CNFET-based cache, we have each cache way to work at its optimized speed. However, the major challenge is to match the variable cache access latency with the actual cache operations that may hit on different cache ways at runtime.

To address the problem, we have a set of delay registers to store the delay of the different cache ways as shown in \autoref{fig:vasa-arch} and the delay registers are then used to adapt the variable cache accesses by enabling the output MUX at the right time. Suppose that the delay of the different cache ways can be determined with testing and configured to the registers at the initialization stage. For each cache request, the request tag is compared to tags of all the different cache ways in parallel to determine the target cache way. Since the data width of the tag is small compared to the width of the cache data, the tag read and comparison is fast. The comparison result is then utilized to select the delay register of the corresponding cache way. When the delay of the target cache way is obtained, a delay controller will issue an enabling signal to acknowledge the output MUX and output the corresponding cache data at the right time. Since the delay controlling path with smaller data width is faster than the cache data access, the delay controlling will not induce additional delay to the cache accesses. In this case, the cache access latency is essentially determined by the access latency of the cache way in which the data is located. Thereby, it outperforms the baseline cache that always operates with the worst cache way latency.

\subsection{Latency-aware Data Shuffling}
To further make use of the CNFET-based cache with varied cache way access latency, we propose a latency-aware data shuffling strategy which has the most recently used data stored in the cache way with shorter latency to achieve higher performance. The proposed data shuffling strategy is built on top of the conventional least recently used (LRU) strategy. Basically, we prioritize the cache data based on their access history and update the priorities at runtime. When a cache data is accessed, it becomes the most recently used cache data and is assigned with the highest priority. The priorities of the rest of the cache data in the same cache set can be affected and need to be updated accordingly. When each cache data in the same cache set is prioritized differently, many data movements are required to match the cache data priority with the corresponding cache access latency. For instance, when the most frequently used cache data is located at the slowest cache way, the entire cache set needs to be updated. Meanwhile, suppose we have 8 cache ways, we need at least 3bit priority for each cache data and each cache set needs to be extended by 24bits, which consumes considerable chip area. In this work, we propose a cache grouping strategy to have each cache set divided into four groups based on their access latency. The priorities are managed with the granularity of a cache group rather than each cache data. In this case, the cache update can be reduced and less priority bits are required. 

To illustrate the proposed data shuffling, we have a representative example presented in Figure \ref{fig:data-shuffling} which includes four typical data shuffling scenarios. Note that the 1-bit 'T’ field represents the cache access priority within a group and lower value indicates higher priority. The color of the blocks represents the access latency of the cache ways and light color stands for lower latency. The four data shuffling scenarios under different cache hits are shown as follows.
\begin{itemize}
    \item In Figure \ref{fig:data-shuffling}(a), cache hits on the fast cache way group (G0). The data block in the cache Way 1 becomes the most recently used data. The priority of the previously most recently used cache data in cache Way 0 is degraded to '1' accordingly. Since the data block is already in the fastest cache way group (G0), there is no need for the data shuffling.
    
    \item In Figure \ref{fig:data-shuffling}(b), cache hits on cache way group G1. The data block in cache Way 3 becomes the most recently used data, so we need to swap it to the fastest cache way group (G0). We place the data block in cache Way 0 which has '1' tag in G0. To conduct the data block swapping in parallel, we have two registers utilized for the intermediate data buffering. During this data shuffling, only cache Way 0 and cache Way 3 are active.
    
    \item In Figure \ref{fig:data-shuffling}(c), cache hits on the slowest cache way group (G3). The data block in cache Way 7 becomes the most recently used data and we place it in cache Way 1 in the fastest cache way group (G0). Nevertheless, it can be a waste if we move the original data block in cache Way 1 to the slowest cache way group (G3) directly, because the data block in cache Way 1 is recently utilized and it is more likely to be reused later. Thereby, we choose to move it to cache Way 3 and move the original data block in cache Way 3 to the next cache way group (G2). This approach smooths the data shuffling. 
    
    \item In Figure \ref{fig:data-shuffling}(d), cache miss occurs and new data block will be placed at cache Way 1. While cache Way 1 stores the data with relatively higher priority, we will not evict the data block in cache Way 1 directly. Similarly, we move the data block with '1' tag (Way 1) to the next group (G1), replace another cache data block with '1' tag block, and invert the priority bit. Eventually, the data block with lower priority (Way 7) in the slowest cache group (G3) turns out to be the victim data block and is kicked out of the cache.
\end{itemize}

%% file: vawa.tex
\section{Variation-aware Way Aligned (VAWA) Cache} \label{sec:vawa}
In this section, we mainly investigate the cache architecture design and optimization for the variation-aware way aligned (VAWA) cache.

\subsection{VAWA Cache Architecture}
Similar to VASA cache, VAWA cache also has a delay controller to synchronize the data read from the cache based on the the access latency of each cache set in the delay registers. Nevertheless, there are a large number of cache sets with different access latency, the number of delay registers can be extremely large compared to that in VASA which requires only a few delay registers to record the access latency of different cache ways. For instance, a 2MB cache with 8 cache ways includes 4096 cache sets and 4096 delay registers are required, which is expensive in terms of hardware implementation. 

\begin{figure}
	\center{\includegraphics[width=0.9\linewidth]{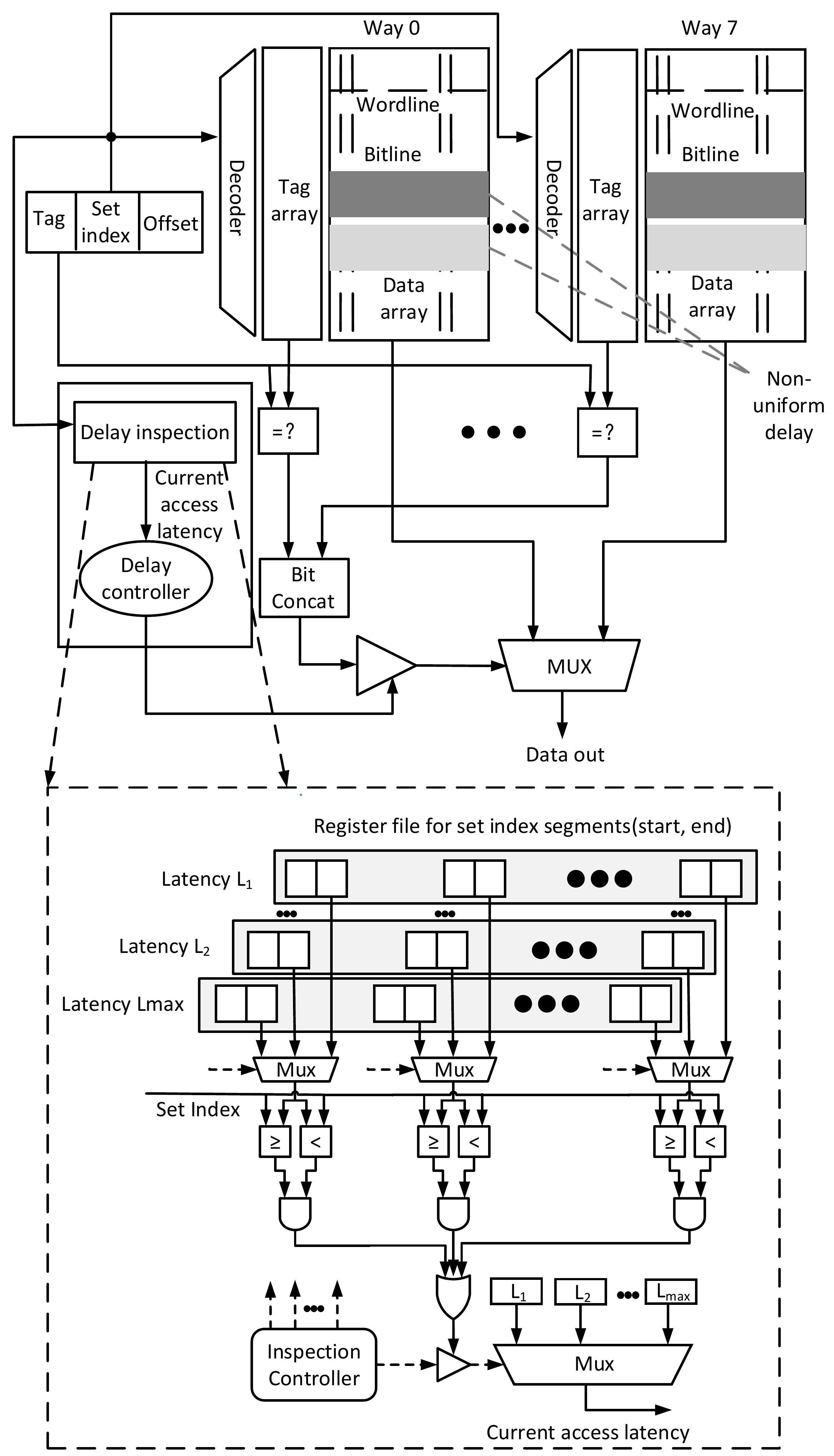}}
	\vspace{-0.5em}
    \caption{Variation-aware way aligned cache architecture.}
\label{fig:vawa}
\vspace{-1em}
\end{figure}

An intuitive approach to address this problem is to divide the cache sets evenly into groups such that cache sets in the same group can share the same delay register and the number of delay registers can be greatly reduced. Nevertheless, the number of cache sets in each group still ranges from dozens to hundreds and the cache access latency of each group is determined by the slowest cache set. While the cache sets with higher access latency can be located in arbitrary locations, the probability that a group includes at least a high-latency cache set remains non-trivial, which limits the cache access latency of the entire group. We notice that the majority of the cache sets actually have very low access latency while only a small number of cache sets suffer high access latency according to the model in \cite{li2015microarchitectural}. With this observation, we propose a non-uniform grouping algorithm to reduce the average cache access latency with a limited number of delay registers. The basic idea is to only record the locations of the low-latency cache sets while leaving the rest of the cache sets working at the worst timing. Suppose the cache set latency varies from $L$ cycle to $H$ cycles. For the cache sets with $M$-cycle latency ($L \leq M \leq H$), we have a set of location pairs $(S_i, E_i)$ to record the starting index and the ending index of the consecutive cache set segments where $i \in [1, N]$ and $N$ indicates the number of cache set segments with $M$-cycle latency. Since the number of consecutive cache set segments with the same latency may be larger than $N$, we only choose the segments with most cache sets. In addition, we may also take $M$ as $\leq M$ because a cache set with less than $M$-cycle latency can also work at $M$-cycle latency. As the non-uniform grouping algorithm can be performed offline and the number of cache sets is limited, we utilize a brute-force implementation to determine the optimized cache set grouping. 

With the grouping, each cache request needs to check if it hits on the low-latency cache groups. If yes, the corresponding cache operation will be adapted to the low latency cache access with the variation-aware architecture. Otherwise, the cache operation will be adapted to the highest access latency. The proposed VAWA architecture is shown in \autoref{fig:vawa}. It can be seen that the basic variation architecture is consistent with that in VASA. A delay controller is used to adjust the output MUX selection to select the cache data with dynamic arrival time at the right cycle. The major difference lies on the latency management. Instead of having the latency of different cache partitions stored in the registers, VAWA mainly stores the locations of the cache sets with the same latency. In VAWA, there are three groups of cache sets. Two of them have low-latency cache sets included and each group has 16 register pairs to keep the indices of the cache set segments. The third group includes all the rest cache sets and works with the worst timing. A cache request will compare with all the indices of cache set segments in the same cache set group in parallel, but the comparison of different cache groups are performed sequentially. Thus, the comparison units can be reused. The lower latency cache group will be compared first such that the comparison will not stall the cache operations. If a cache request does not belong to the first two groups, it must belong to the last group accordingly, so there is no need to have the corresponding cache set indices stored in the registers.

\subsection{Variation-aware Page Mapping}
\begin{figure}
	\center{\includegraphics[width=0.95\linewidth]{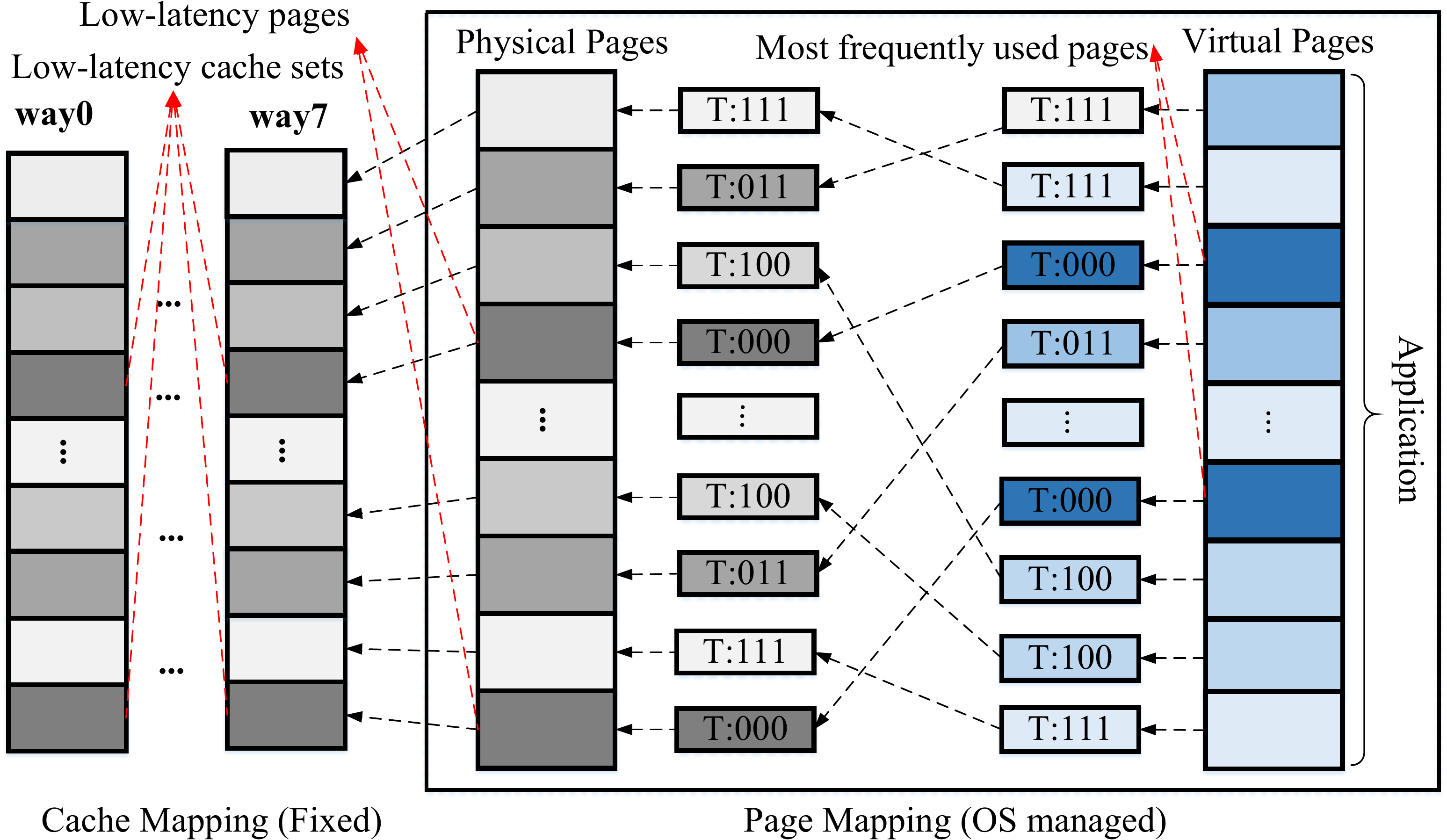}}
	\vspace{-0.5em}
    \caption{Variation-aware page mapping.}
\label{fig:page-mapping}
\vspace{-1em}
\end{figure}

In order to take advantage of the delay variations across the different cache sets, we propose a variation-aware page mapping approach such that the frequently used data can be mapped to the cache sets with shorter latency. As shown in Figure \ref{fig:page-mapping}, the first step of this approach is to identify the frequently used data through memory access profiling. While the cache set index that a data is mapped to is typically determined by the physical address of the data, we choose to revise the page mapping that converts the virtual addresses to physical addresses such that OS can map the frequently accessed virtual pages to an available physical page that will be loaded to the faster cache sets. 

Basically, we have the virtual pages classified into multiple levels based on the virtual page utilization. Frequently used pages are tagged with lower levels.  At the same time, as physical addresses mapped to cache sets are usually known and fixed, the cache access latency of each physical address can also be determined. In this case, we can also have the physical addresses tagged based on the potential cache access latency. When a virtual page is converted to a physical page, the level information of the virtual pages will also be used to determine the physical page allocation. Specifically, the virtual pages with lower levels will be mapped to physical pages with low-latency tags accordingly. In addition, as the mapping is managed with the granularity of pages rather than per cache line, it also poses constraints to the cache grouping. Suppose the page size is $P$ KB, the cache grouping must be conducted with the granularity of $G$ cache sets where $G=P/(L \times W)$, $W$ refers to the number of cache ways, and $L$ refers to the cache line size such that each page can be mapped to consecutive cache sets with the same access latency. Particularly, the non-uniform grouping must have each segment configured with at least $G$ cache sets. The page mapping step is done in the operating system during the page translation from virtual addresses to physical addresses, so it does not require any additional hardware modifications to the cache design. 

%% file: NUCA.tex
\section{Variation-aware Non-Uniform Cache Architecture (NUCA)} \label{sec:NUCA}
\begin{figure}
	\center{\includegraphics[width=0.65\linewidth]{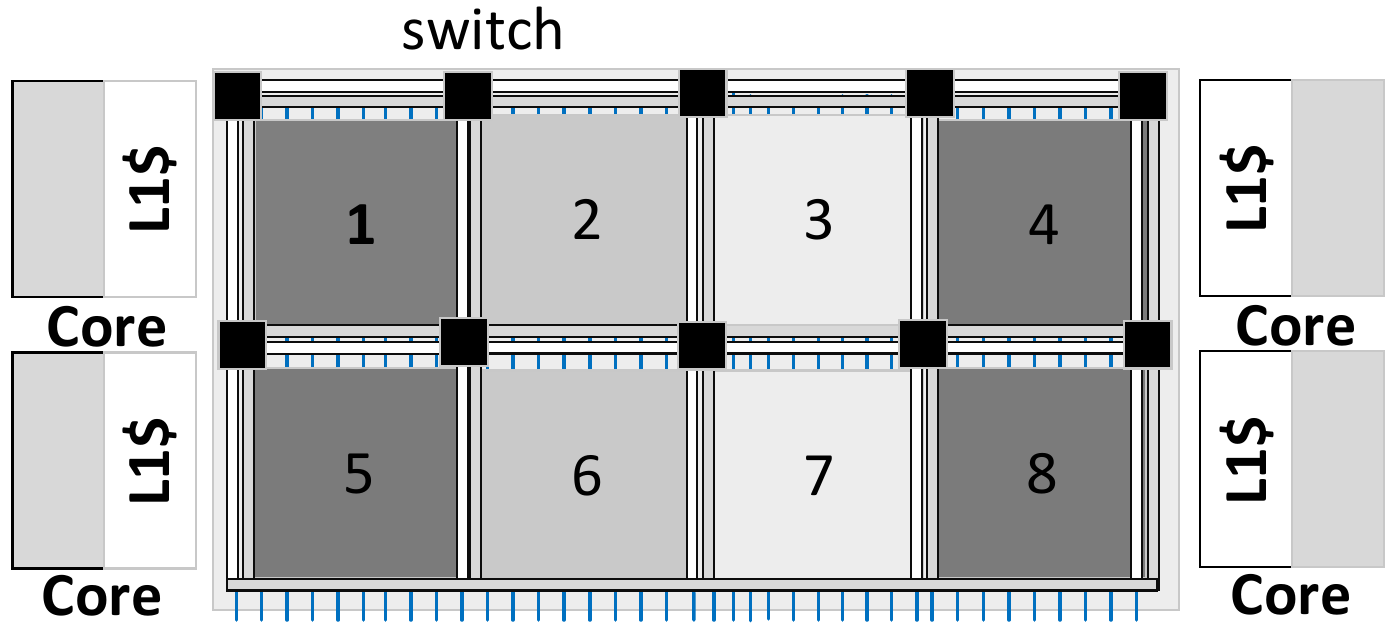}}
	\vspace{-0.5em}
    \caption{Baseline CNFET-based NUCA. It includes 8 cache banks and each bank is a CNFET-based variation-aware cache unit.}
\label{fig:NUCA-arch}
\vspace{-1em}
\end{figure}

Both VAWA and VASA target at a uniform cache architecture and they can not be deployed on a non-uniform cache architecture (NUCA) directly which has inherent latency variations across the different cache banks. A baseline CNFET-based NUCA example is shown in Figure \ref{fig:NUCA-arch}. It includes multiple cache banks connected with a regular 2D mesh Network-on-Chip (NoC) architecture. Each bank of the distributed cache is implemented with the CNFET-based cache architecture. As a result, the cache access latency depends on both the cache bank locations and the CNT growth direction, which complicates the latency variation problem and makes the CNFET-based NUCA design challenging. To address the above problem, we take both the cache bank location incurred access latency variations and the CNFET PV induced access latency variations into consideration, and combine the two latency variations into a unified model such that we can estimate NUCA access latency details and unleash the potential of the CNFET-based NUCA with os-level page mapping which allocates the frequently used data to the cache banks with lower latency \cite{managing2006cho}. 

The latency model is key to the NUCA optimization and it roughly includes two parts as illustrated in Equation \ref{eq:latency}. The first part represents the average access latency of a cache bank. It mainly characterizes the latency variations caused by the CNFET PV and depends on the underlying CNT layouts as discussed in prior sections. When the CNT growth direction is parallel to the bitline, the proposed VASA cache structure is utilized for each cache bank of the NUCA. Nevertheless, the proposed data shuffling optimization for VASA can only be applied to a single cache bank due to the communication overhead across the different cache banks. In this case, we utilize the average cache way latency as the $hit\_lat$ for each different cache bank. Given the same CNFET PVs, the average hit latency for the different cache banks are supposed to be the same in theory. Hence, we mainly take the VASA NUCA as a normal NUCA. When the CNT growth direction is parallel to the wordline, each cache bank is implemented as VAWA architecture. While each cache bank can be viewed as multiple cache groups with different access latency. From the perspective of page mapping, there is no difference between cache groups in different cache banks and cache groups in the same cache bank. Hence, the entire NUCA can be viewed as a unified VAWA with multiple cache groups of different access latency. The access latency of each cache group can also be represented by Equation \ref{eq:latency}. 

The second part of the cache access latency represents the latency variations induced by the spatial locations of the distributed cache banks. It depends on the distance between the cache bank and the CPU core that issues the cache access. They can be roughly estimated with the average NoC hops and calculated with the corresponding cache bank index and the core index in NoC. Particularly, the requests from the different cores can have distinct access latency to the same cache bank. In summary, the cache hit latency is generally independent with the CNFET PVs and it can be considered as a latency offset added to the different cache banks while the communication latency mainly depends on the relative distance between the cores and the target cache banks. NUCA with VAWA-based cache banks can be optimized with a unified page mapping with the granularity of cache groups across the NUCA. Although NUCA with VASA-based cache banks can also be optimized with the page mapping across the different cache banks, the internal data shuffling is only applicable within each cache bank because the communication overhead between the different cache banks hinders the adoption of the fine-grained data shuffling.

\begin{equation}
\label{eq:latency}
    Lat = Hit\_Lat + NoC\_Lat(Core\_ID, Bank\_ID)
\end{equation}

%% file: experiment.tex
\section{Experiments} \label{sec:experiment}
\subsection{Experiment Setup}
We modified GEM5 \cite{binkert2011gem5} to implement and evaluate the CNFET-based LLC in a CMP architecture. The CNFET technology parameters are presented in Table \ref{tab:cnt} and the CMP architectural parameters are listed in Table \ref{tab:cmp}. We have a UCA and a NUCA implemented as LLC i.e. L2 in CMP respectively. Particularly, NUCA is divided into 8 cache banks and these banks are interconnected with a $2 \times 4$ 2-D mesh NoC. Four CPU cores are attached to the routers in the corner of NoC and each cache bank is attached to a router in the NoC. Single-cycle routers with X-Y routing is 
applied for the NoC. We selected four single-thread programs including gcc, mcf, gromac and libq from SPEC CPU2006 and four parallel programs including fft, LU, nonLU and radix from splash2 \cite{woo1995splash} as the benchmark. In order to estimate the energy consumption of the CNFET-based cache design, we obtained the energy consumption of a basic SRAM cell with CNFET SPICE model and scales it to the entire cache \cite{zhang2012carbon}. For the NoC-based NUCA, we only have the SRAM-based cache banks evaluated as we do not modify the NoC for the NUCA.

\newcommand{\tabincell}[2]{\begin{tabular}{@{}#1@{}}#2\end{tabular}} 
\begin{table}
    \centering
  \caption{CMP architectural parameters}
  \label{tab:cmp}
  \begin{tabular}{cc}
    \toprule  
      Paremeters& Setting  \\
    \midrule
      Number of core & 4  \\
      Core frequency & 2GHZ \\
      Conformance protocol & MESI \\
      Cache replacement strategy & LRU \\
      Cache line size & 64B \\
      Memory access latency & 30 cycles \\
      L1 icache & \tabincell{c}{16Kb, 2-way associative,\\64 byte  per block, 1 cycle} \\
      L1 dcache & \tabincell{c}{32Kb, 2-way associative,\\64 byte per block, 1 cycle} \\
      L2 cache & \tabincell{c}{2Mb, 8-way associative,\\64 byte per block, 6-12 cycle} \\
      
  \bottomrule
\end{tabular}
\vspace{-1em}
\end{table}


In this work, we evaluated the variation-aware UCA and NUCA architectures with different CNFET layouts from the perspective of both performance and energy consumption. There are two typical layouts i.e. set aligned cache and way aligned cache for the CNFET-based cache architectures. Hence, four major design categories including VAWA UCA, VASA UCA, VASA NUCA, and VAWA NUCA need to be evaluated. At the same time, we have a baseline cache, which is operated at the worst timing to ensure unified cache hit latency to the entire cache, implemented for comparison. On top of the baseline cache, we have a classic CNFET-based circuit optimization approach applied \cite{li2015microarchitectural}. The basic idea of this approach is to disable the cache ways with the worst timing such that the entire cache can work at higher clock but smaller capacity. It provides a design trade-off between the cache operation speed and the cache capacity. The partial disabling approach is abbreviated as baseline+PD and it is also implemented for comparison. 

\subsection{Uniform Cache Architecture (UCA) Evaluation}
\subsubsection{VASA UCA}
For the set aligned UCA, the access latency of different cache ways range from 6 to 12 cycles and the exact access latency of each cache way is determined with the model \cite{li2015microarchitectural}. The baseline cache access latency is limited to the cache ways with the worst timing i.e. 12-cycle. In contrast, the proposed VASA architecture can have the cache ways with latency variations working at their optimized timing. Moreover, in order to further take advantage of the access latency variations in each cache line, we take the data reuse frequency into consideration and have a latency-aware data shuffling (DS) approach implemented in VASA to swap the frequently used data to cache ways (within the same case set) with lower latency at runtime. This approach is abbreviated as VASA+DS. 

\begin{figure}
	\center{\includegraphics[width=0.95\linewidth]{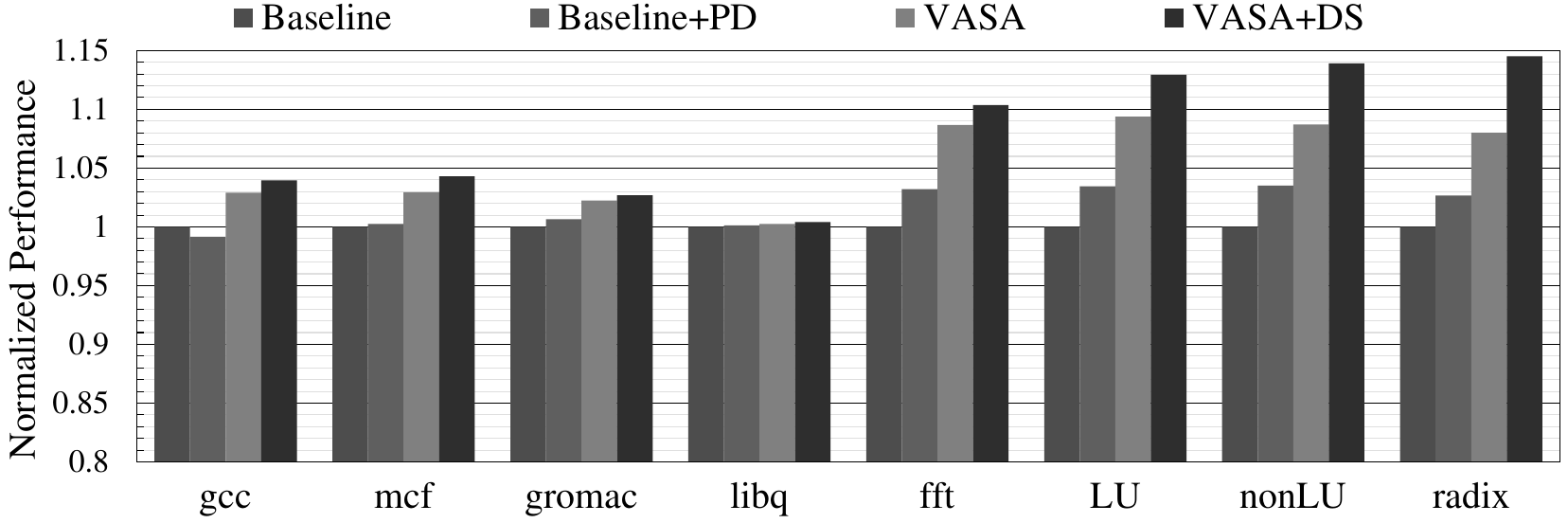}}
    \caption{Normalized benchmark performance on processors with set aligned CNFET LLC.}
\label{fig:UCA-VASA-performance}
\end{figure}

\textbf{Performance:}
We have the benchmark performance on processors with baseline, baseline+PD, VASA, VASA+DS UCA compared and the comparison result is shown in \autoref{fig:UCA-VASA-performance}. Baseline refers to the cache working at the worst timing i.e. 12-cycle. Baseline+PD refers to the baseline cache with partial disabling optimization which disables the cache ways with the worst timing. It can be seen that VASA cache outperforms the baseline cache notably on all the benchmark programs while VASA+DS cache with DS shows additional performance improvement on top of the VASA cache. Although Baseline+PD that lowers the average cache access latency with some capacity penalty shows moderate performance improvement relative to the baseline on most of the benchmark programs, it remains less competitive to the proposed VASA and VASA+DS. At the same time, we notice that VASA cache made the major contribution to the performance improvement while DS also poses non-trivial contribution. In general, VASA+DS shows 6\% performance improvement on average compared to the baseline, but we find that VASA UCA achieves significantly higher performance improvement on parallel benchmark programs than the single-thread benchmark programs.

\begin{figure}
	\center{\includegraphics[width=0.95\linewidth]{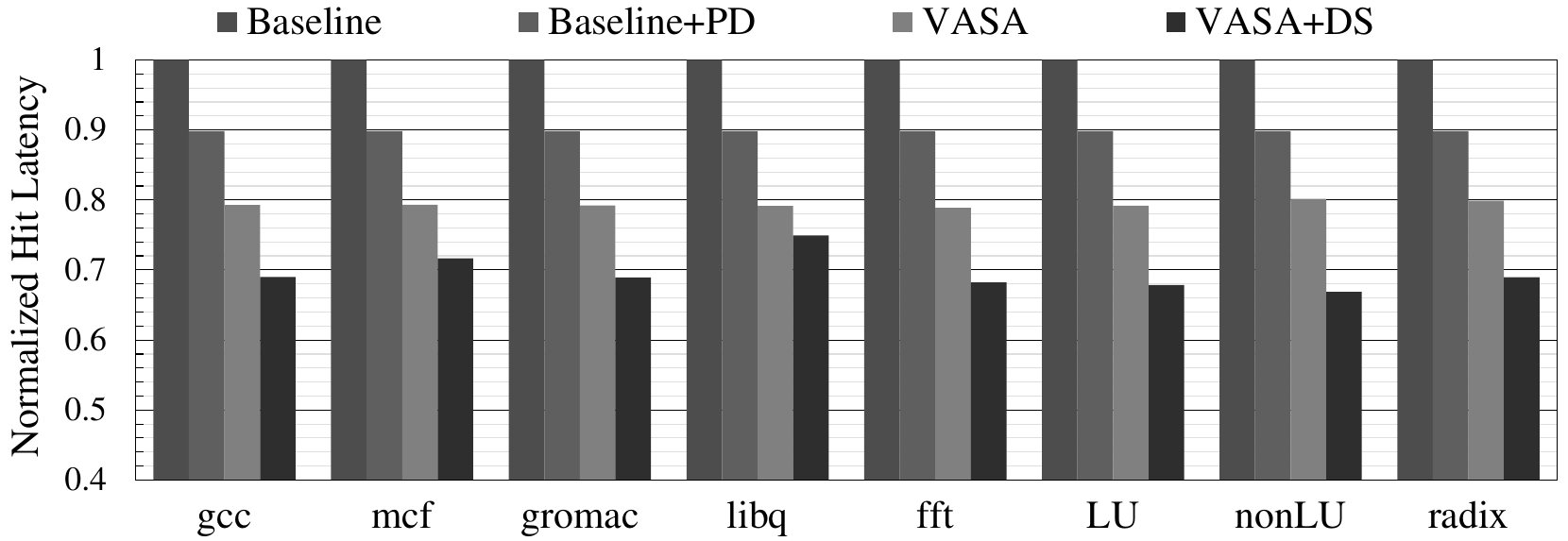}}
    \caption{Normalized cache hit latency of the set aligned CNFET LLC.}
\label{fig:UCA-VASA-latency}
\vspace{-1em}
\end{figure}

To further investigate the performance comparison, we analyzed the cache miss rate as shown in \autoref{tab:vasa} and the average cache hit latency as presented in \autoref{fig:UCA-VASA-latency}. It can be found that the average cache hit latency of VASA+DS reduces by around 32\% on average compared to the baseline cache. The average cache hit latency mainly depends on the underlying hardware, it shows little variation across the different benchmark programs. On the other hand, we find that L1 icache miss rate on parallel benchmark programs is much higher. Since the L1 cache miss eventually leads to LLC cache access and icache is more sensitive to the cache access latency, the same LLC cache hit latency reduction poses more significant influence on the performance of the parallel benchmark programs, which explains the performance improvement difference of the benchmark programs. In contrast, libq from Spec CPU2006 is a single-thread memory-bound task and suffers 99.23\% LLC cache misses. There are very few LLC cache hit that can benefit from the proposed DS optimization. Thereby, libq performance improvement induced by the LLC cache hit latency reduction is trivial.

\begin{table}
    \centering
  \caption{Cache miss rate of VASA UCA$^a$.}
  \label{tab:vasa}
  \begin{tabular}{cccc}
    \toprule  
     benchmark & L1 icache miss &  L1 dcache miss & L2 cache miss \\
     \midrule
     gcc & 0.0348 &	0.0424 & 0.3507 \\
     mcf & 0.0012 &	0.3537 & 0.6497 \\
     gromac & 0.0068 & 0.1006 & 0.0369 \\
     libq & 0.000006 & 0.3050 & 0.9923 \\
     fft & 0.1506 & 0.0101 & 0.0649 \\
     LU & 0.1518 & 0.0032 & 0.0022 \\
     nonLU & 0.1606 & 0.0087 & 0.0022 \\
     radix & 0.2434 & 0.0026 & 0.0037 \\
  \bottomrule
\end{tabular}
\footnotesize{\\ $^a$ We also have the cache miss rate analyzed under cache configurations including VAWA UCA, VASA NUCA, and VAWA NUCA. The cache miss rate that mainly depends on the cache sizes and the inherent locality of the program shows little difference. In this case, the cache miss rate in this table will be used for analysis of different cache configurations to save the space}
\vspace{-1em}
\end{table}

\begin{figure}
	\center{\includegraphics[width=0.95\linewidth]{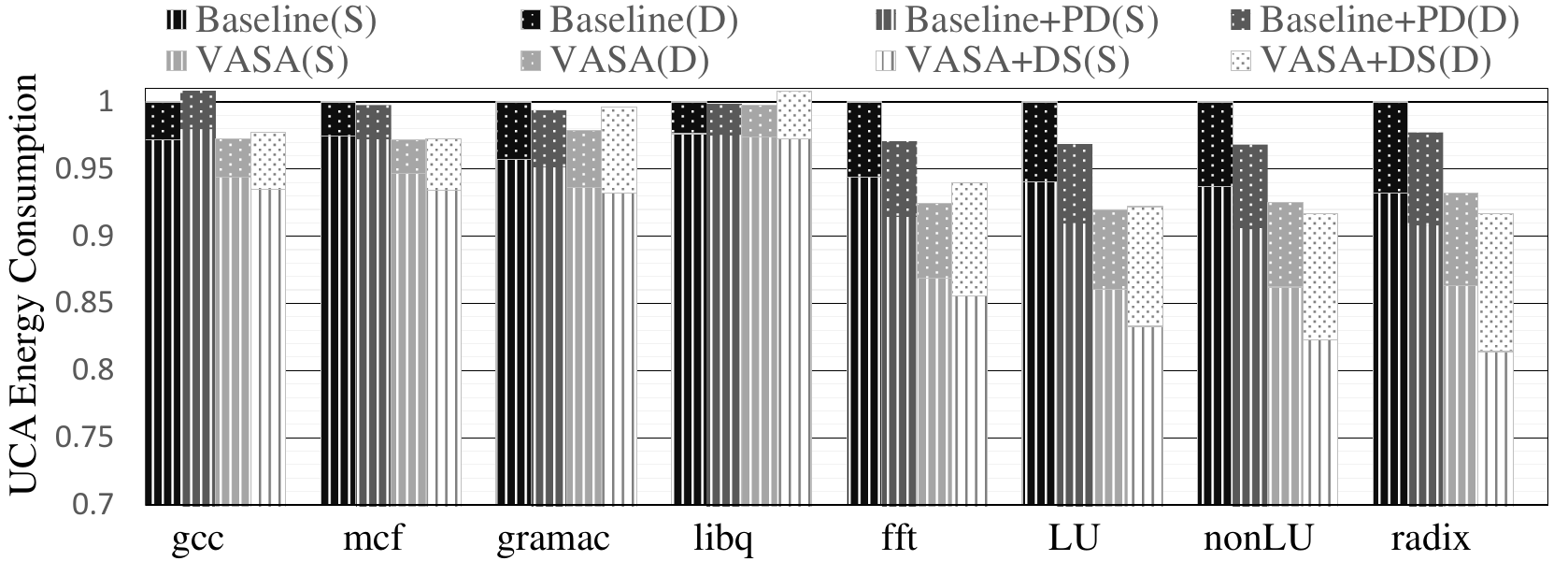}}
	\vspace{-0.5em}
    \caption{Normalized energy consumption of different CNFET LLC implementations with set aligned layout.(S stands for static energy, D stands for dynamic energy.)}
\label{fig:VASA-energy}
\vspace{-1em}
\end{figure}

\textbf{Energy Consumption:} We investigate the energy consumption of the different cache implementations when the benchmark is executed. The cache energy consumption mainly consists of dynamic energy consumption and static energy consumption. The static energy consumption is the production of the program runtime and the static power consumption. Since the static power consumption changes little with the workloads, the static energy consumption is mainly determined by the program runtime. The dynamic energy consumption is the production of per cache operation energy and the number of cache operations required in each program. While per cache operation energy mainly depends on the cache architecture and the cache size, it varies little for a specific cache configuration. Hence, the dynamic energy consumption mainly depends on the number of cache operations in the programs eventually. The normalized energy consumption of the cache implementations under different benchmark programs is illustrated in \autoref{fig:VASA-energy}. It can be seen that the static energy consumption dominates the LLC energy consumption in general because the majority of the cache hits on L1 cache and the LLC is inactive in most cases. In this case, the total energy consumption is mostly consistent with the benchmark runtime. Nevertheless, DS induces additional cache write to swap frequently used data to cache ways with lower latency, which also incurs non-trivial energy consumption. When the energy overhead cannot be compensated by the performance improvement induced energy saving, the total energy consumption can be even higher. This is observed on gcc, gramac, fft, and LU according to \autoref{fig:VASA-energy}. Still, VASA+DS achieves 4\% energy saving on average compared to the baseline.

\textbf{Hardware Overhead} Due to the lack of CNFET technology library, we cannot provide the chip area of the cache implementations directly. While the hardware overhead of the proposed cache optimizations is mainly attributed to the extended cache cells and the register files, the overhead of the rest logic added to the cache is trivial and ignored. For VASA UCA , it only requires 4 bytes register files and extra logic control circuit. For VASA + DS , it requires 8 bits extended to each cache line for the runtime data access analysis which takes up 4096 bytes in total. At the same time, it needs 260 bytes register files to be used for the data shuffling and the cache way latency recording, which is small compared to the cache sizes.

     





\subsubsection{VAWA UCA}
For the way aligned CNFET cache, we propose a variation-aware way aligned (VAWA) cache accordingly to make use of the cache sets with different access latency. The access latency varies from 6 to 10 cycles and the exact access latency of each cache set can be determined with the model \cite{li2015microarchitectural}. While a straightforward look up table based latency aware mechanism needs tremendous registers to store the different latency configurations of the cache sets, we divide the cache sets into limited number of groups and each group has a uniform access latency. Two different cache grouping algorithms including a uniform grouping and a non-uniform grouping are implemented. For the uniform grouping, we split the cache sets evenly into 64 groups. For the non-uniform grouping, we divide the cache sets into three groups depending on its latency. The first group contains the cache sets with 6-cycle latency and the second group contains the cache sets with 7-cycle latency. There are 128 registers used to record the locations of the cache sets for two cache groups. The third group contains all the rest cache sets and they are operated at the worst timing i.e. 10-cycle. In order to take advantage of the cache groups with different access latency, we further propose a latency-aware page mapping strategy to assign frequently used data to faster cache groups. As the cache grouping and the page mapping is closely coupled, we abbreviate the page mapping with different grouping algorithms as VAWA+UG+PM and VAWA+NG+PM respectively. 

\begin{figure}
	\center{\includegraphics[width=1\linewidth]{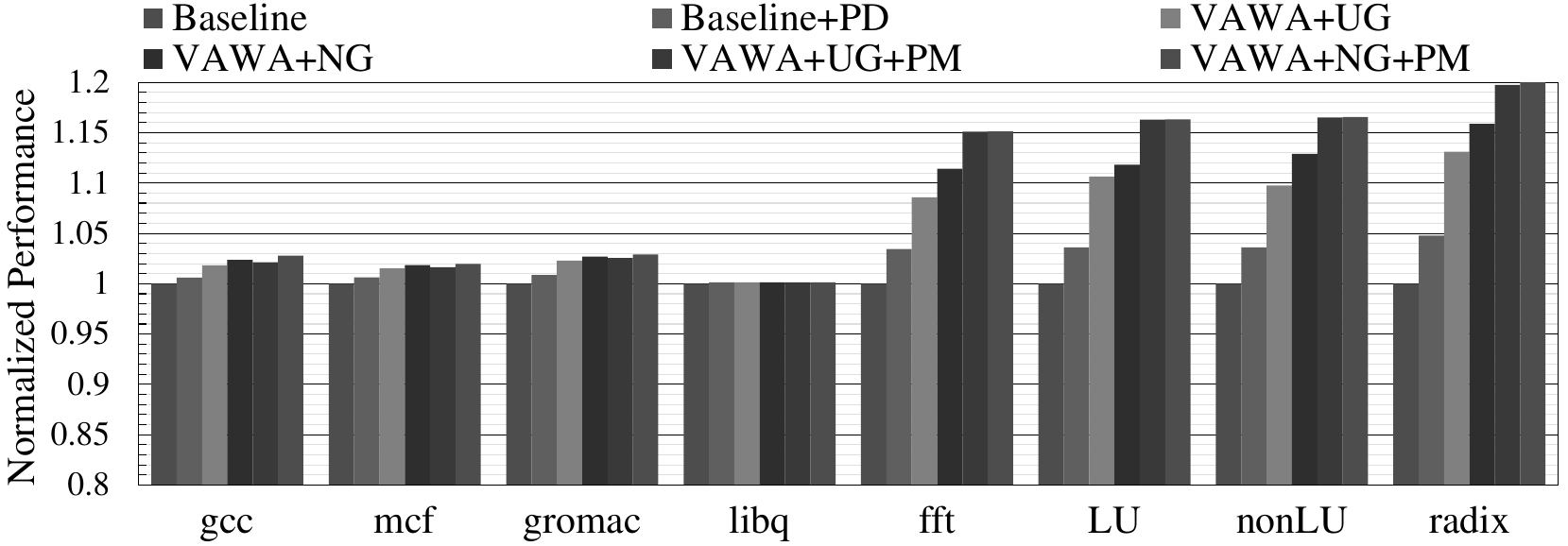}}
	\vspace{-2em}
    \caption{Normalized benchmark performance on processors with way aligned CNFET LLC.}
\label{fig:UCA-VAWA-performance}
\vspace{-1em}
\end{figure}

\begin{figure}
	\center{\includegraphics[width=1\linewidth]{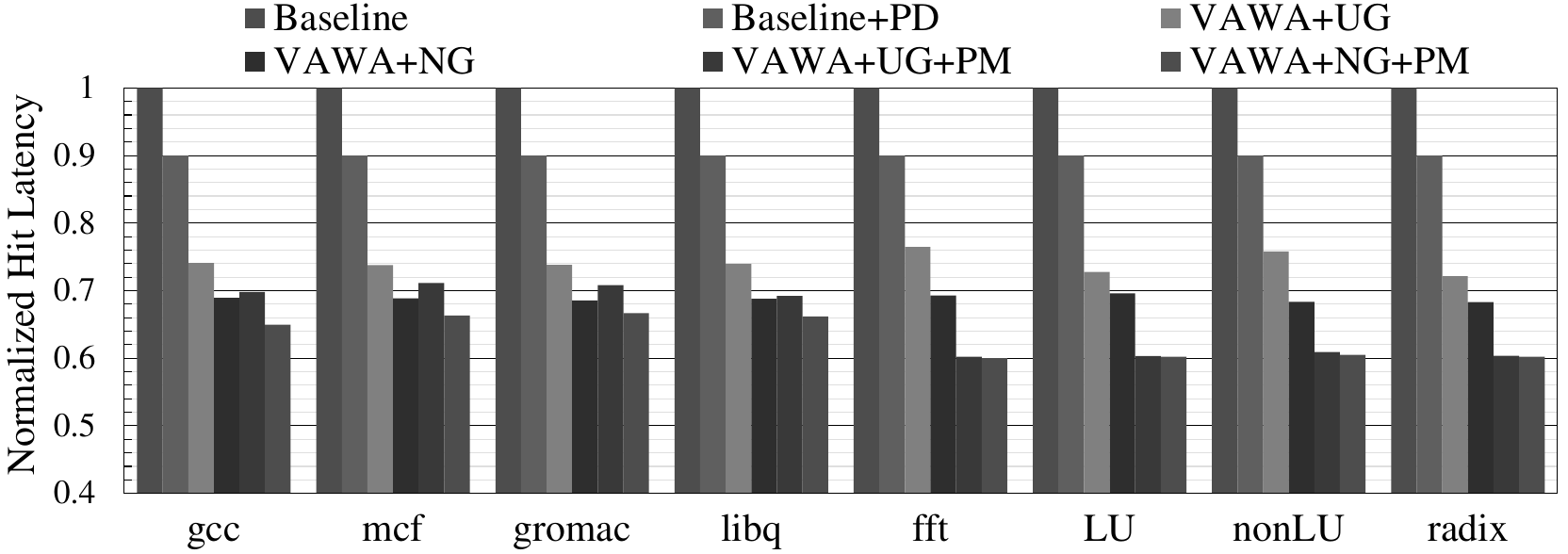}}
	\vspace{-2em}
    \caption{Normalized cache hit latency of the way aligned CNFET LLC.}
\label{fig:UCA-VAWA-latency}
\vspace{-1em}
\end{figure}

\textbf{Performance:} For the way aligned CNFET cache, we have a set of different cache architectures including baseline, baseline+PD, VAWA+UG, VAWA+NG, VAWA+UG+PM, VAWA+NG+PM implemented and compared. Similar to set aligned cache, baseline refers to the cache working at the worst timing i.e. 10-cycle. Baseline+PD refers to the baseline cache with partial disabling optimization which disables the cache sets with the worst timing. VAWA+UG and VAWA+NG refer to the proposed variation-aware CNFET cache architecture with uniform cache grouping and non-uniform cache grouping respectively. The performance of all the cache architectures is compared in \autoref{fig:UCA-VAWA-performance}. VAWA+NG+PM outperforms the other designs in general and achieves 9\% performance improvement on average over the baseline. At the same time, it can be seen that the variation-aware architecture contributes most to the performance improvement while the page mapping also exhibits non-trivial contribution in general. The proposed non-uniform cache grouping also shows moderate performance improvement over the uniform cache grouping when there is no page mapping. However, the advantage of non-uniform grouping to uniform grouping is greatly undermined particularly on the parallel benchmark programs when the page mapping optimization is applied. The main reason is that these parallel programs typically need LLC to fetch instructions rather than data because they show relatively high L1 icache miss but lower L1 dcache miss according to \autoref{tab:vasa}. The instructions usually have better spatial locality and a small portion of low-latency cache sets can meet the requirements. Specifically, we analyzed the LLC hits of the parallel benchmark programs and found that more than 95\% cache hits are located at around 1\% cache sets. In this case, non-uniform grouping that offers more low-latency cache sets poses little influence on the performance of the parallel benchmark programs. According to the cache access latency in \autoref{fig:UCA-VAWA-latency}, VAWA+NG+PM shows almost no cache access latency reduction compared to VAWA+UG+PM. which confirms the above analysis.

As shown in \autoref{fig:UCA-VAWA-latency}, although the average cache hit latency reduction is roughly consistent with the performance improvement, the cache access latency reduction amplitude is much larger than that of the performance improvement. Specifically, the average cache hit latency of VAWA+NG+PM is 30-40\% lower than that of the baseline while the performance improvement is less than 5\% on the single-thread benchmark programs and around 16\% on parallel benchmark programs. Particularly, we notice that the LLC access latency reduction on libq fails to promote its performance. There are mainly two reasons. First, latency of the data accesses to the LLC can be partly hidden by the instruction scheduling in the processors. Thus, data accesses are less sensitive to the access latency while instruction accesses are more sensitive and have more significant influence on the performance. Second, the number of LLC cache accesses varies across the different programs. More performance improvement can be expected from the access latency reduction when there are more intensive LLC accesses issued. Parallel benchmark programs with much higher L1 icache miss rate mainly rely on the LLC for the instruction fetch, so it is more sensitive to the latency reduction and exhibits higher performance improvement. libq has low L1 dcache miss rate but very high LLC cache miss rate according to \autoref{tab:vasa}. It indicates that there are only very few cache hit on LLC. Thus, the cache hit latency contributes little to libq performance. 

\begin{figure}
	\center{\includegraphics[width=0.98\linewidth]{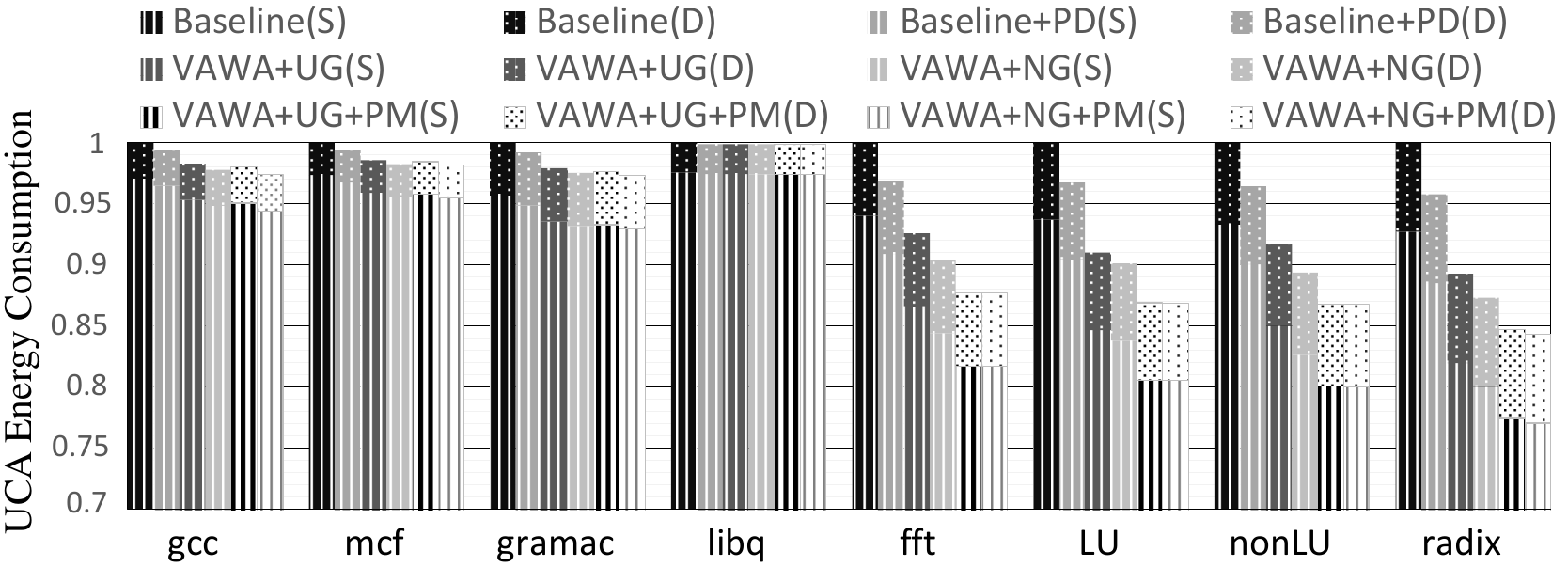}}
	\vspace{-0.5em}
    \caption{Normalized energy consumption of different CNFET LLC implementations with way aligned layout.(S stands for static energy, D stands for dynamic energy.)}
\label{fig:VAWA-energy}
\vspace{-1em}
\end{figure}

\textbf{Energy Consumption:} The energy consumption of the way aligned cache with different optimizations is presented in \autoref{fig:VAWA-energy}. VAWA+NG+PM consumes the least energy in general and it is 8\% lower than the baseline on average. Unlike VASA cache that induces more dynamic energy consumption, VAWA-based optimizations do not need to extend the cache way sizes nor additional cache operations, so the dynamic energy consumption remains similar across the different optimizations and the energy saving is mainly attributed to the performance improvement.  


\textbf{Hardware Overhead:}
For the way aligned cache, the added hardware overhead for VAWA+UG and VAWA+NG is mainly attributed to the registers that are used to store the latency information of different cache groups and the locations of cache sets with specific latency respectively. The register size for VAWA+UG is 224-byte and the register size for VAWA+NG is 194-byte. Page mapping does not need to modify the cache design, so it will not induce additional hardware overhead.



\subsection{NUCA Evaluation}
In this subsection, we mainly evaluate the CNFET-based NUCA with two typical layouts i.e. set aligned layout and way aligned layout, and focus on the performance and energy consumption evaluation. For the different layouts, we have VASA cache architecture and VAWA cache architecture used directly for each cache bank in NUCA accordingly. 

\subsubsection{VASA NUCA}
\begin{figure}
	\center{\includegraphics[width=0.95\linewidth]{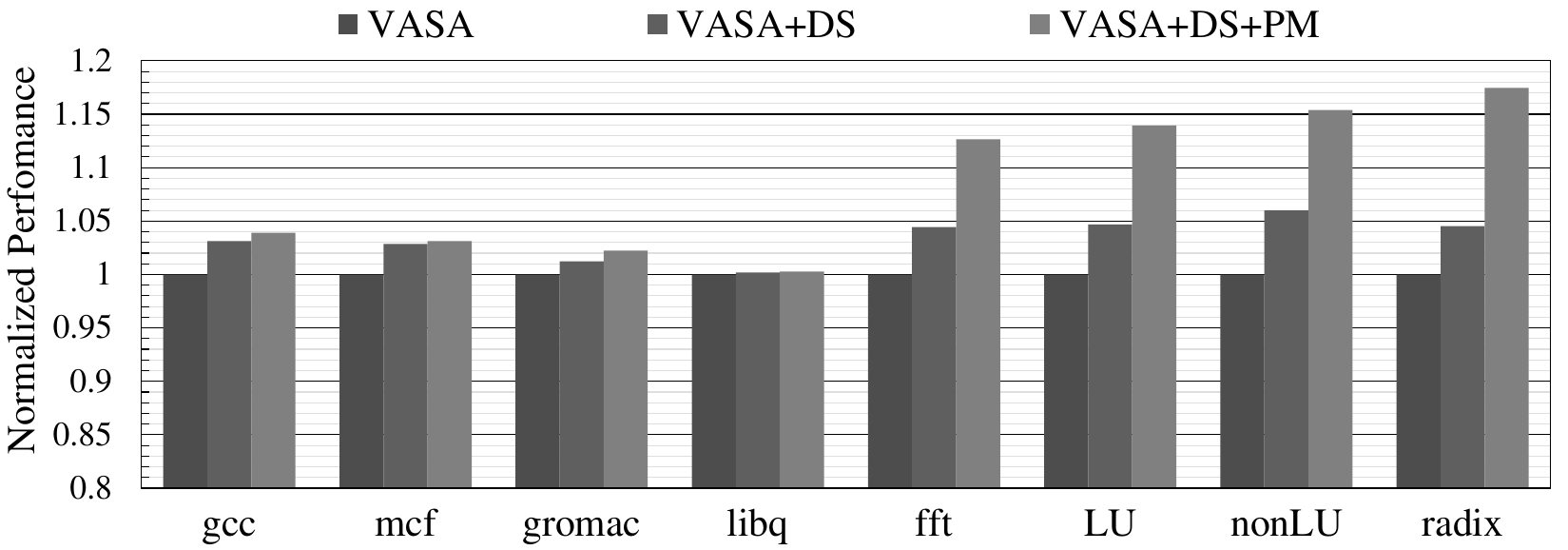}}
	\vspace{-0.5em}
    \caption{Normalized performance of different CNFET NUCA implementations with set aligned layout.}
\label{fig:NUCA-VASA-performance}
\vspace{-1em}
\end{figure}

\begin{figure}
	\center{\includegraphics[width=0.95\linewidth]{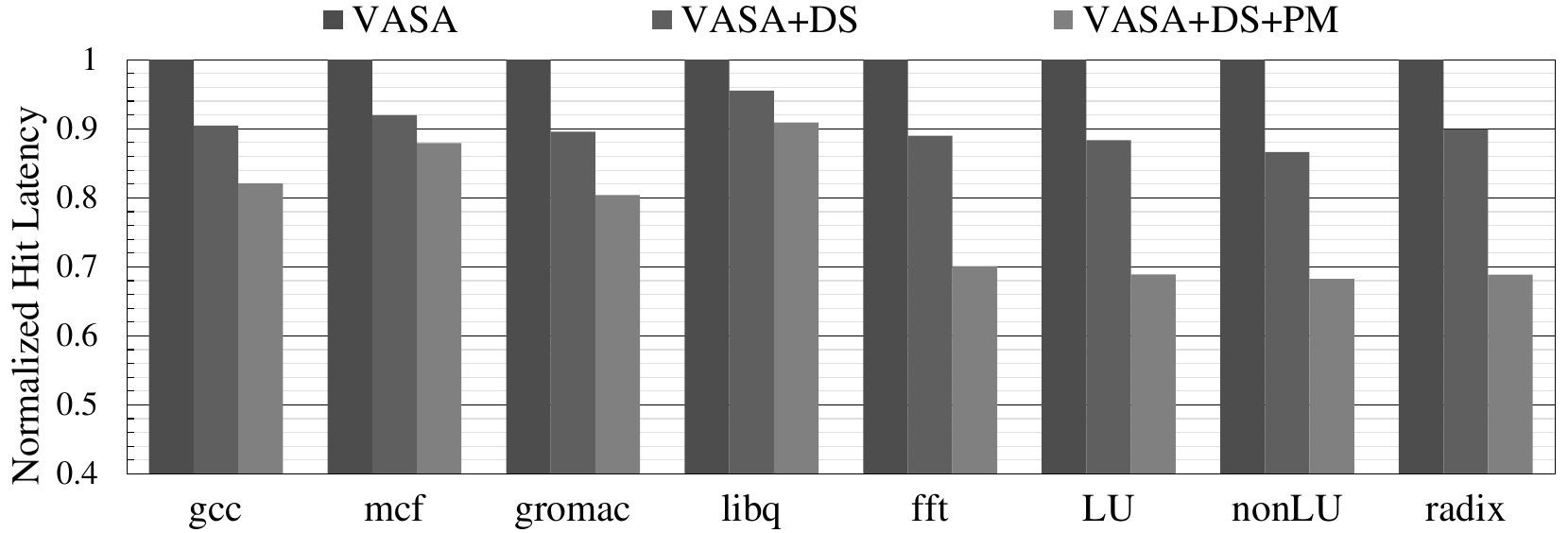}}
    \caption{Normalized cache hit latency of different CNFET NUCA implementations with set aligned layout.}
\label{fig:NUCA-VASA-latency}
\vspace{-1em}
\end{figure}

\textbf{Performance:}
Unlike UCA, NUCA suffers latency variations inherently due to the distributed cache banks spread across a 2-D mesh NoC despite the underlying CNFET layouts. Thus, page mapping is applied to both VASA NUCA and VAWA NUCA to make use of the cache access variations. For VASA NUCA, the page mapping is conducted with the granularity of a cache bank and we utilize the average access latency of all the cache ways in a cache bank as the metric. This approach is abbreviated as VASA+PM. In addition, we also have a basic NUCA that works with the VASA cache banks and VASA+DS cache banks without page mapping implemented and compared. The comparison is presented in \autoref{fig:NUCA-VASA-performance}. It can be seen that VASA+DS+PM shows significant higher performance than the other independent optimizations on all the benchmark programs, which indicates that both the optimizations are necessary and they can be combined. Particularly, parallel benchmark programs with considerable instruction accesses to LLC are more sensitive to the access latency and exhibit more significant performance improvement. The average cache hit latency presented in \autoref{fig:NUCA-VASA-latency} confirms the benefits brought by the cache hit latency reduction. At the same time, we notice that the page mapping is critical to address the cache bank spatial location induced latency variations and it shows substantial access latency reduction compared to the straightforward VASA with data shuffling particularly for parallel benchmark programs.

\begin{figure}
	\center{\includegraphics[width=0.95\linewidth]{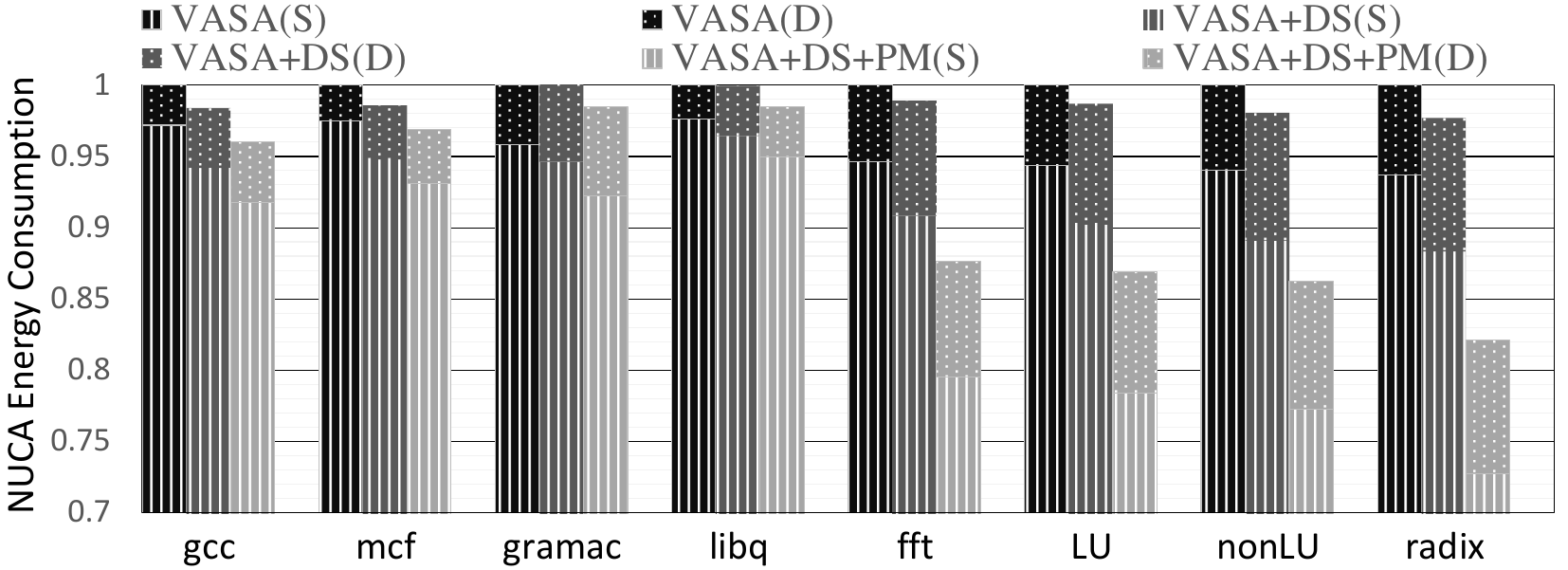}}
	\vspace{-0.5em}
    \caption{Normalized energy consumption of CNFET NUCA with set aligned layout.(S stands for static energy, D stands for dynamic energy.)}
\label{fig:NUCA-VASA-energy}
\vspace{-1em}
\end{figure}

\textbf{Energy Consumption:}
\autoref{fig:NUCA-VASA-energy} reveals the energy consumption of the proposed NUCA implementations with set aligned layout. Generally, VASA+DS+PM shows lowest energy consumption on all the benchmark programs while the energy reduction is much higher on parallel benchmark programs because of the significant higher performance speedup. In contrast, we notice that the data shuffling requires additional cache writes, so the data shuffling induces higher dynamic power consumption, which is consistent with the VASA+DS UCA.

\subsubsection{VAWA NUCA}
\begin{figure}
	\center{\includegraphics[width=0.95\linewidth]{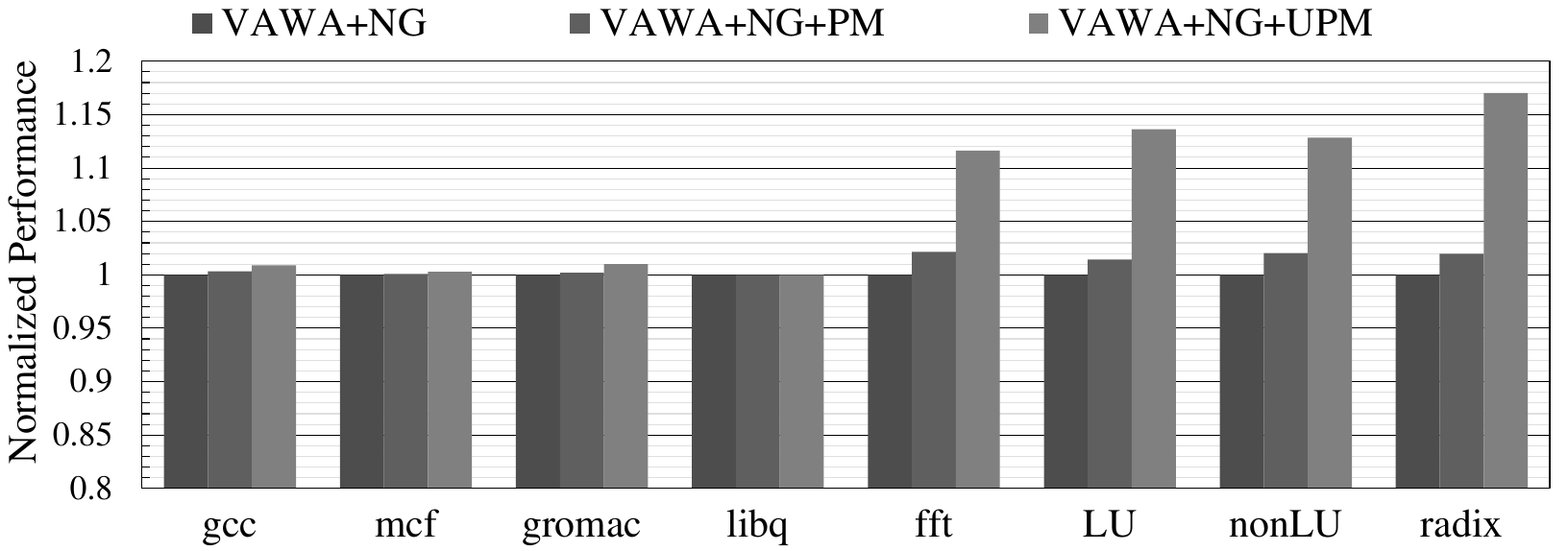}}
	\vspace{-0.5em}
    \caption{Normalized performance of different CNFET NUCA implementations with way aligned layout.}
\label{fig:NUCA-VAWA-performance}
\vspace{-1em}
\end{figure}

\begin{figure}
	\center{\includegraphics[width=0.95\linewidth]{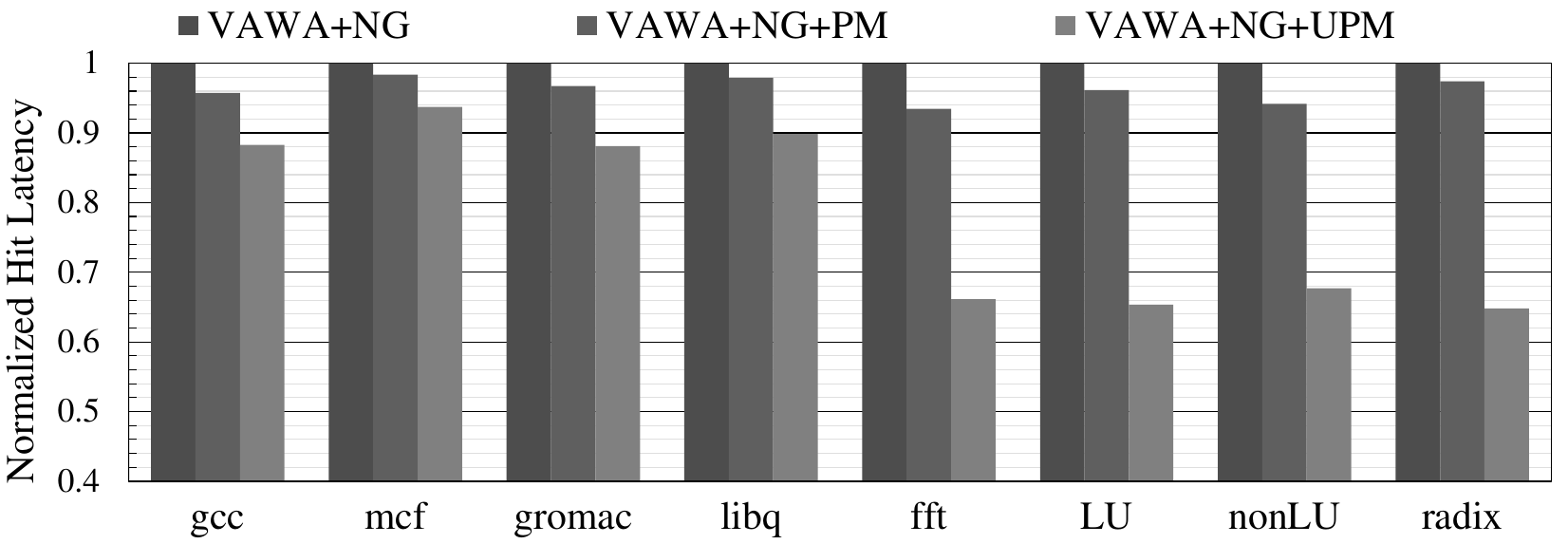}}
	\vspace{-0.5em}
    \caption{Normalized cache hit latency of different CNFET NUCA implementations with way aligned layout.}
\label{fig:NUCA-VAWA-latency}
\vspace{-1em}
\end{figure}

\textbf{Performance:} For VAWA NUCA, we have a unified latency variation model to characterize the different latency variations and exploit the potential of the cache with unified page mapping. This implementation is abbreviated as VAWA+NG+UPM. At the same time, we have two different NUCAs implemented for comparison. One of them is constructed with the basic VAWA cache banks and each cache bank is optimized with non-uniform grouping. It is denoted as VAWA+NG. The other NUCA implementation has page mapping applied to all the VAWA+NG cache banks without considering their spatial locations. This implementation is denoted as VAWA+NG+PM. The performance comparison of the three way aligned NUCA implementations is shown in \autoref{fig:NUCA-VAWA-performance}. VAWA+NG+UPM shows 
notable performance improvement compared to both VAWA+NG and VAWA+NG+PM especially on parallel benchmark programs. It indicates that page mapping is a key factor to address the access latency variations caused by both CNFET PVs and cache spatial locations. Meanwhile, we notice that the average performance speedup is more significant than that on VAWA UCA. This is mainly because that NUCA enlarges the cache access latency variations and benefits more from the page mapping optimization. This is also demonstrated by the average cache hit latency comparison presented in \autoref{fig:NUCA-VAWA-latency}.

\begin{figure}
	\center{\includegraphics[width=0.95\linewidth]{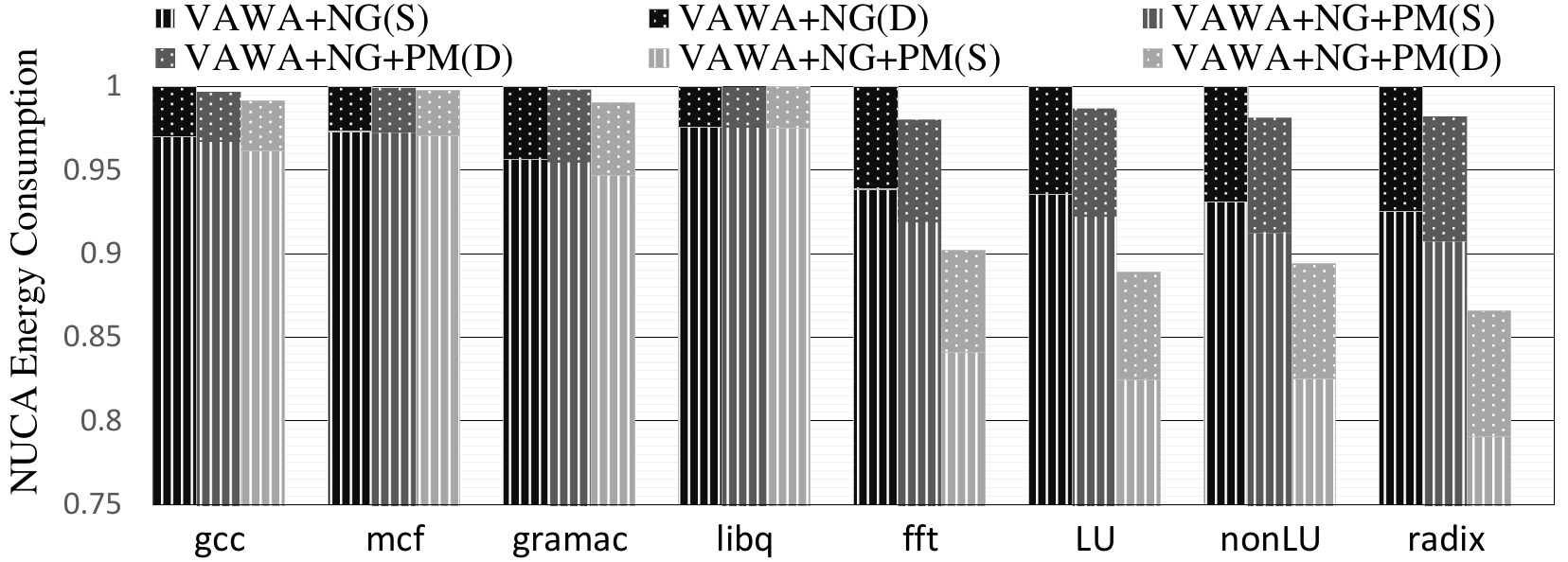}}
	\vspace{-0.5em}
    \caption{Normalized energy consumption of different CNFET NUCA implementations with way aligned layout.(S stands for static energy, D stands for dynamic energy.)}
\label{fig:NUCA-VAWA-energy}
\vspace{-1em}
\end{figure}

\textbf{Energy Consumption:}
The energy comparison of different cache implementations is presented in \autoref{fig:NUCA-VAWA-energy}. Generally, the page mapping optimization will not induce additional hardware overhead nor energy consumption, so it has little influence on the dynamic energy consumption. In contrast, the static energy consumption is reduced considerably because of the runtime reduction as presented in \autoref{fig:NUCA-VAWA-performance} particularly on multi-thread benchmark programs.

